\documentclass[a4paper,fleqn,usenatbib]{mnras}

\usepackage{newtxtext,newtxmath}

\usepackage[T1]{fontenc}
\usepackage{ae,aecompl}

\usepackage{graphicx}	
\usepackage{amsmath}	
\usepackage{amssymb}	
\usepackage{multicol}        
\usepackage{bm}		
\usepackage{pdflscape}	

\title[Metallicity Gradients of the Thick Disc Progenitor]{Metallicity Gradient of the Thick Disc Progenitor at High Redshift}

\author[D. Kawata et al.]{
Daisuke Kawata,$^{1}$\thanks{E-mail: d.kawata@ucl.ca.uk}
Carlos Allende Prieto,$^{2}$
Chris B. Brook,$^{2}$
Luca Casagrande,$^{3}$
\newauthor{Ioana Ciuc${\rm \breve{a}}$,$^{1}$
Brad K. Gibson,$^{4,9}$
Robert J. J. Grand,$^{5,6}$  Michael R. Hayden,$^{7}$}
\newauthor{and Jason A. S. Hunt$^{8}$}
\\
$^{1}$Mullard Space Science Laboratory, University College London, Holmbury St. Mary, Dorking, Surrey, RH5 6NT, UK\\
$^{2}$Instituto de Astrof\'{i}sica de Canarias, V\'{i}a L\'{a}ctea, 38205 La Laguna, Tenerife, Spain\\
$^{3}$Research School of Astronomy \& Astrophysics, Mount Stromlo Observatory, Australian National University, ACT 2611, Australia\\
$^{4}$E.A. Milne Centre for Astrophysics, University of Hull, Hull, HU6 7RX, United Kingdom \\
$^{5}$Heidelberger Institut f\"{u}r Theoretische Studien, Schloss-Wolfsbrunnenweg 35, 69118 Heidelberg, Germany\\
$^{6}$Zentrum f\"{u}r Astronomie der Universit\"{a}t Heidelberg, Astronomisches Recheninstitut, M\"{o}nchhofstr., 69120 Heidelberg, Germany\\
$^{7}$Laboratoire Lagrange (UMR7293), Universite de Nice Sophia Antipolis, CNRS, Observatoire de la Cote d'Azur, BP 4229, F-06304\\ Nice Cedex 4, France\\
$^{8}$Dunlap Institute for Astronomy and Astrophysics, University of Toronto, Ontario M5S 3H4, Canada\\
$^{9}$Joint Institute for Nuclear Astrophysics, Center for the Evolution of the Elements (JINA-CEE)\\
}

\date{Accepted XXX. Received YYY; in original form ZZZ}

\pubyear{2017}

\begin{document}
\label{firstpage}
\pagerange{\pageref{firstpage}--\pageref{lastpage}}
\maketitle

\begin{abstract}
We have developed a novel Markov Chain Mote Carlo (MCMC) chemical "painting" technique to explore possible radial and vertical metallicity gradients for the thick disc progenitor. In our analysis we match an N-body simulation to the data from the Apache Point Observatory Galactic Evolution Experiment (APOGEE) survey. We assume that the thick disc has a constant scale-height and has completed its formation at an early epoch, after which time radial mixing of its stars has taken place. Under these assumptions, we find that the initial radial metallicity gradient of the thick disc progenitor should not be negative, but either flat or even positive, to explain the current negative vertical metallicity gradient of the thick disc. Our study suggests that the thick disc was built-up in an inside-out and upside-down fashion, and older, smaller and thicker populations are more metal poor. In this case, star forming discs at different epochs of the thick disc formation are allowed to have different radial metallicity gradients, including a negative one, which helps to explain a variety of slopes observed in high redshift disc galaxies. This scenario helps to explain the positive slope of the metallicity-rotation velocity relation observed for the Galactic thick disc. On the other hand, radial mixing flattens the slope of an existing gradient.
\end{abstract}




\section{Introduction}
\label{sec:intro}

The Milky Way's thick disc was originally identified from the analysis of star counts as a function of the vertical height \citep{yy82,gilr83}. This traditional definition of the thick disc corresponds to a geometrically-defined thick disc \citep[e.g.][]{rc86,yy92,rm93,jibls08}. On the other hand, high-resolution spectroscopic studies of the solar neighbourhood stars show two sequences in the  $\alpha$-element to the iron abundance ratio, [$\alpha$/Fe], as a function of the iron abundance, [Fe/H] \citep[e.g.][]{kf98,pncmw00,fbl03}. The higher [$\alpha$/Fe] sequences is usually considered as a chemically-defined thick disc. The chemically defined thick disc could be different from the geometrically-defined thick disc \citep{mmssdjs15} \citep[see also][for a discussion about the various definitions of thick and thin discs]{dkcc16}. It is now confirmed that a similar high-[$\alpha$/Fe] sequence exists over a large radial and vertical range of the Galactic disc \citep[e.g.][]{nbbah14, hbhnb15}. From the Gaia-ESO survey, \citet{mhrbdv14} analysed the radial metallicity gradients of the chemically-defined thin and thick disc populations within the region at $6<R<12$ kpc and $|z|<0.6$ kpc. They found that the chemically-defined thick disc shows no radial metallicity gradient \citep[see also][]{apbwnryl06}, but a shallow negative vertical metallicity gradient. This indicates that the chemical properties of the chemically-defined thick disc are well mixed radially, but maintain a negative vertical gradient \citep[see also][]{baboym11,nbbah14,hbhnb15}. Comparing the spectroscopically derived stellar parameters with theoretical isochrones, \citet{hdmlkg13} analysed the stellar ages for chemically-defined thick and thin discs. In general, the chemically-defined thick disc stars are older than the chemically-defined thin disc stars \citep[see also][]{brsfa14,mg15}. Although some young high-[$\alpha$/Fe] disc stars are also found \citep{carmm15,mrahm15}, some of them are likely to be blue stragglers \citep[e.g.][]{jjvimhg16,ycvckmm16} and their origin is still debated. Therefore, in this paper, we consider that the chemically-defined thick disc stars are generally old, and define our thick disc as the chemically-defined old thick disc.
 
 
 Advanced analysis of spectroscopic data starts to reveal the radial metallicity gradients of disc galaxies at $z\sim1-3$, which is likely the formation epoch for the thick disc. \citet{yksrl11} found a very steep radial metallicity gradient of $-0.16\pm0.02$ dex kpc$^{-1}$ for a lensed disc galaxy at $z=1.49$. \citet{jerj13} also reported two out of four lensed rotation-dominated galaxies observed at $z\sim2$ showing radial metallicity gradients steeper than $-0.2$~dex~kpc$^{-1}$. If the thick disc of the Milky Way had this steep metallicity gradient, there must be some mechanism to flatten this gradient (without heating the thin disc component) in order to explain the current lack of a radial gradient in the Milky Way's thick disc. 
 
 Interestingly, observations of more disc galaxies at high redshift  show a variety of the gradients, including a significant fraction of galaxies with a flat metallicity gradient \citep{ssgbmct12,ljesrza16,wwffsg16}, and some positive ones \citep[e.g.][]{cmmmgm10}. \citet{wwffsg16} analysed radial  gradients of 180 KMOS$^{\rm 3D}$ near-infrared integral-field-unit (IFU) data at $z=0.6-2.7$, and found that most of their galaxies show flat radial metallicity gradient within uncertainties. Only 15 out of 180 galaxies show a radial metallicity gradient that is significantly negative (13 galaxies) or positive (2 galaxies). 
 
 The evolution of radial metallicity gradients should provide strong constraints on disc formation scenarios \citep[e.g.][]{gpbsb13}, and therefore it is important to compare gradients of different age populations of the Milky Way with Milky Way-like disc galaxies at different redshifts. However, it is not straightforward to infer initial metallicity distribution for mono-age populations at the time they formed, from the current Galactic metallicity distribution as a function of age, because radial mixing can alter the distributions \citep[e.g.][]{sb09a,mcm14,gkc15,kggchb17}. Following \citet{kggchb17}, we refer to ``radial mixing" to describe the overall radial re-distribution due to both ``churning" and ``blurring" \citep{sb09a,rspm17}. Churning indicates the change of angular momentum of stars, and blurring describes the radial re-distribution of stars due to their epicyclic motion. Churning can be split into two mechanisms, ``radial scattering" and ``co-rotation radial migration". Radial scattering describes the change of angular momentum with kinematic heating, i.e. the increase in random energy of the orbit. This includes heating from mergers and satellite bombardment \citep[e.g.][]{gtjpo92,jbkw12}. On the other hand, co-rotation radial migration indicates a gain or loss of angular momentum of stars at the co-rotation resonance of the spiral arms \citep{jsjb02} or bar, and does not involve significant kinematic heating.
 
As mentioned, the Milky Way's thick disc has a flat radial metallicity gradient. The main question addressed in this paper is, what range of radial metallicity gradients is plausible for the progenitor of the Milky Way's thick disc when it formed at high redshift. We use an idealised N-body simulation model for the thick and thin discs, where the thick disc formed at an early epoch, 8~Gyr ago, and experienced radial mixing due to bar and spiral arms developed in the thin disc. To limit the complexity of the simulation and highlight the effects of radial mixing, our numerical experiment does not include the growth of the thin disc. We also assume a constant scale-height for the thick disc, independent of radius, for simplicity. Within these assumptions, we introduce a novel Markov Chain Monte Carlo (MCMC) painting technique which assigns metallicity tags to N-body particles by exploring the possible radial and vertical metallicity gradients before they suffered from radial mixing, attempting to fit the current metallicity distribution of the Milky Way's thick disc observed by the Apache Point Observatory Galactic Evolution Experiment (APOGEE) at different radii and vertical heights \citep{hbhnb15}. 

The formation of the thick disc is expected to have had been completed at an early epoch, before the thin disc starts growing, as suggested by cosmological simulations \citep[e.g.][]{bkgf04b,bkmg06,bsgkh12,bkwgcmm13,sbrbr13,mcm13,mmssdjs15,gsgmp16}. These simulations also suggest that the thick disc formed in an inside-out and upside-down fashion, i.e. the early thick disc was smaller and thicker \citep[e.g.][]{bkmg06,bsgkh12,bkwgcmm13,mmssdjs15}. Using two components for the thick discs, a thicker, smaller one and a thinner, larger one, we also qualitatively study how the inside-out and upside-down formation of the thick disc affects the possible metallicity distribution of the thick disc progenitor. Section~\ref{sec:meth} describes briefly the numerical simulation code and numerical models. In Section~\ref{sec:res}, we present our results. A summary and discussion of this study is presented in Section~\ref{sec:sum}.

\section{Method and Models}
\label{sec:meth}

We use our original Tree N-body code, {\tt GCD+} \citep{kg03a,kogbc13} for the numerical experiments presented in this paper, and model the evolution of a barred disc galaxy similar in size to the Milky Way. We initially set up an isolated disc galaxy which consists of stellar discs, with no bulge component, in a static dark matter halo potential \citep{rk12,gkc12a,kggchb17}. A live dark matter halo can respond to the disc particles by exchanging angular momentum. However, if not properly modelled, a live dark matter halo may introduce some numerical scattering and heating \citep[e.g.][]{dvh13}. In the interest of performing a more controlled experiment and reducing the computational burden, we use a static dark matter halo.
 We use the standard Navarro-Frenk-White (NFW) dark matter halo density profile \citep{nfw97}, assuming  a $\Lambda$-dominated cold dark matter ($\Lambda$CDM) cosmological model with cosmological parameters of $\Omega_0=0.266=1-\Omega_{\Lambda}$, $\Omega_{\rm b}=0.044$, and $H_0=71{\rm kms^{-1}Mpc^{-1}}$:

\begin{equation}
\rho _\text{dm}=\frac{3H_{0}^{2}}{8\pi G}\frac{\Omega _{0}-\Omega_b}{\Omega_0}\frac{\delta _{c}}{cx(1+cx)^{2}},
\end{equation}
where
\begin{equation}
c=\frac{r_{200}}{r_\text{s}}, \;\; x=\frac{r}{r_{200}},
\end{equation}
and
\begin{equation}
r_{200}=1.63\times 10^{-2}\left(\frac{M_{200}}{h^{-1}M_{\odot }}\right)^{\frac{1}{3}} h^{-1}\textup{kpc},
\end{equation}
\noindent where $\delta _{c}$ is the characteristic density of the profile \citep{nfw97}, $r$ is the distance from the centre of the halo and $r_{s}$ is the scale radius. The total halo mass is set to be $M_{200}=1.2\times 10^{12}M_{\odot }$ and the concentration parameter is set at $c=14$. The halo mass is roughly consistent with  recent estimates of the mass of the Milky Way \citep[e.g.][]{pjm11,bhg16}. 

The stellar disc is assumed to follow an exponential surface radial density profile and $\textup{sech}^{2}$ vertical density profile:
\begin{equation}
\rho _\text{d}=\frac{M_\text{d}}{4\pi z_\text{d}R_\text{d}^2}\textup{sech}^{2}\left(\frac{z}{z_\text{d}}\right)\exp \left(-{\frac{R}{R_\text{d}}}\right),
\end{equation}
\noindent where $M _\text{d}$ is the stellar disc mass, $R_\text{d}$ is the scale length and $z_\text{d}$ is the scale height. In this paper, we consider two thick disc components and one thin disk component. The model parameters are summarised in Table~\ref{tab:model}. The first thick disc component (thick1) is smaller and thicker than the 2nd thick disc component (thick2). The rationale behind these two thick disc components is explained later. In Table~\ref{tab:model}, N$_\text{p}$ is the number of particles for each component. This model has 10,000 M$_{\odot}$ for each particle. We apply the spline softening suggested by \citet{pm07}. We set a softening length of 341 pc (the equivalent Plummer softening length is about 113 pc). To relax the initial system, we ran a simulation for 2 Gyr, moving randomly selected particles azimuthally, which suppresses the development of non-axisymmetric structures \citep{pjmwd07}. We then run the simulations for 8 Gyr which roughly corresponds to the age of the thin disc of the Milky Way. Note that the time of $t=0$ in this paper indicates the starting time of this 8 Gyr simulation using the the end-product of the relaxing run as initial conditions. The N-body simulation is similar to the simulation in \citet{kggchb17} where we discuss that numerical heating is negligible in this simulation setting.

The simulation of the 8~Gyr evolution of the system mimics the evolution history of the Galactic disc after the thick disc formed. For simplicity, we set a thin disc already fully grown from the start, ignoring its growth. This is too simplistic, but we are interested in the effect of radial mixing driven by the thin disc on the thick disc stars. To focus on radial mixing driven by non-axisymmetric structures in the thin disc, we do not include mergers or tidal interaction with the sub-halos or satellite galaxies. This assumption could be justified for the Milky Way, since there is ample observational evidences that the Milky Way has had a quiescent accretion history \citep[e.g.][]{rrfpb14,dbkggmp17}. If there was a merger which was violent enough to induce a strong phase mixing of disc stars after they formed, we expect that it would be difficult to maintain the observed negative metallicity gradient of the thick disc and mono-age populations of the Galactic disc \citep[see also][]{xlyhw15,cklcsc17}, because strong mixing would erase the gradient.

Two component thick discs allow us to explore the inside-out and upside-down formation scenario of the thick disc \citep[e.g.][]{bkgf04b,bkwgcmm13}. We set the thick1 disc being heavier than the thick2 disc. This mimics a scenario that the thick1 disc has grown before the thick2 disc formed, and the thick2 disc represents the latest generation of the thick disc. We study what radial and vertical metallicity gradients these two thick disc components should "originally" have had by comparing with the APOGEE observations of metallicity distribution function (MDF) at different Galactocentric radii and height \citep{hbhnb15}.

 \begin{figure*}
	\includegraphics[width=0.8\hsize]{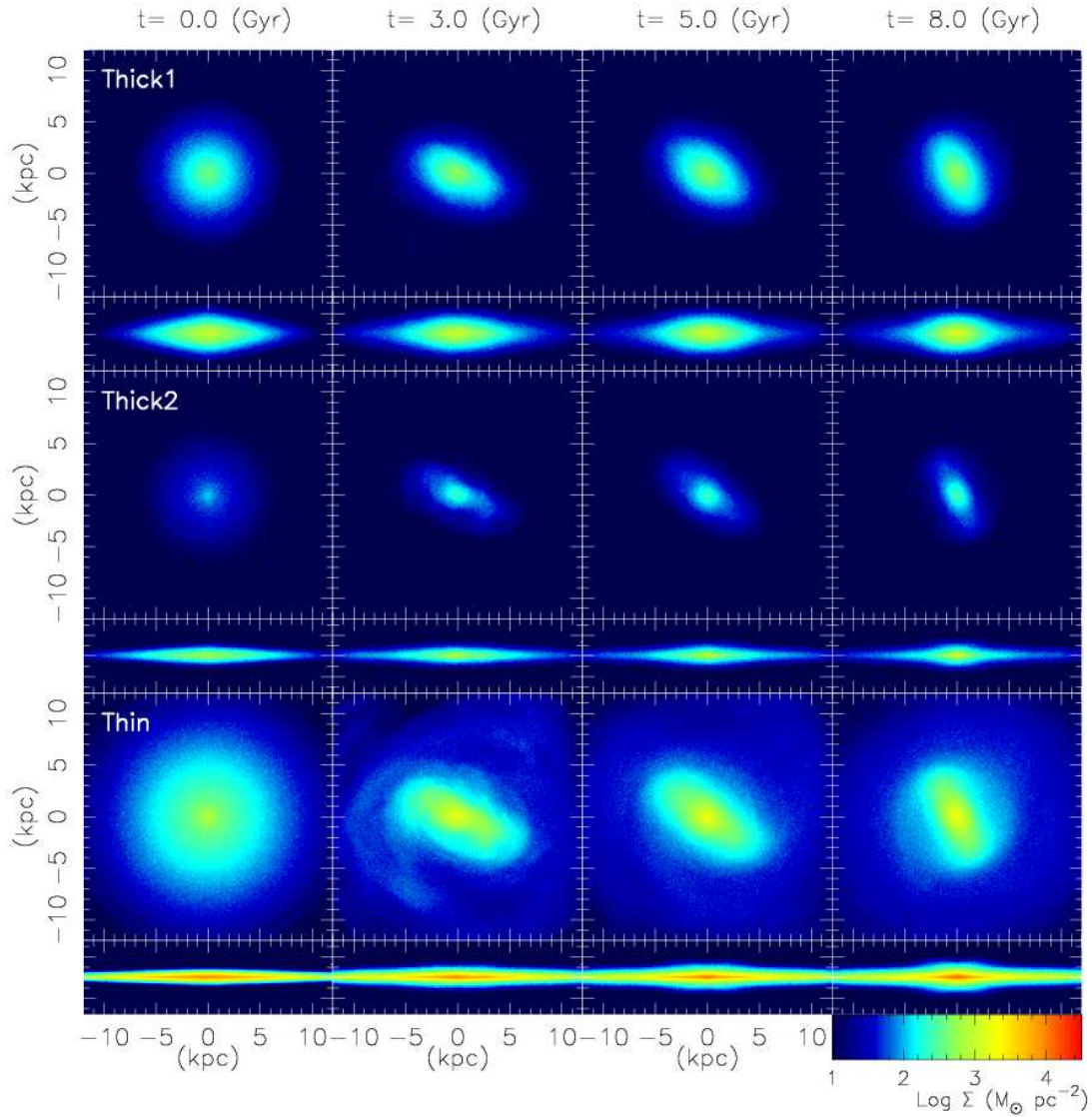}
    \caption{Snapshots of our N-body model which show the the face-on (top, 3rd and 5th panels) and edge-on (2nd, 4th and bottom panels) images of the thick1, thick2 and thin disc components, respectively. }
    \label{fig:snap}
\end{figure*}

\begin{table*}
 \caption{N-body model parameters.}
 \label{tab:model}
 \begin{tabular}{lccccccccccccc}
  \hline
  \multicolumn{4}{c}{Thick1} & & \multicolumn{4}{c}{Thick2} & & \multicolumn{4}{c}{Thin} \\
  \cline{1-4} \cline{6-9} \cline{11-14}
   $M_\text{d,thick1}$ & $R_\text{d,thick1}$ & $z_\text{d,thick1}$ & N$_\text{p,thick1}$ & & $M_\text{d,thick2}$ & $R_\text{d,thick2}$ & $z_\text{d,thick2}$ & N$_\text{p,thick2}$ & & $M_\text{d,thin}$ & $R_\text{d,thin}$  & $z_\text{d,thin}$ & N$_\text{p,thin}$ \\
   $M_{\sun}$ & kpc & kpc & & & $M_{\sun}$ & kpc & kpc & & & $M_{\sun}$ & kpc & kpc &  \\
  \hline
$1.0\times10^{10}$ & 2.0 & 1.0 & 1,000,000 & & $5.0\times10^{9}$ & 3.0 & 0.5 & 500,000 & & 
 $4.5\times10^{10}$ & 3.5 & 0.25 & 4,500,000 \\
  \hline
 \end{tabular}
\end{table*}

\begin{figure}
  \includegraphics[width=\hsize]{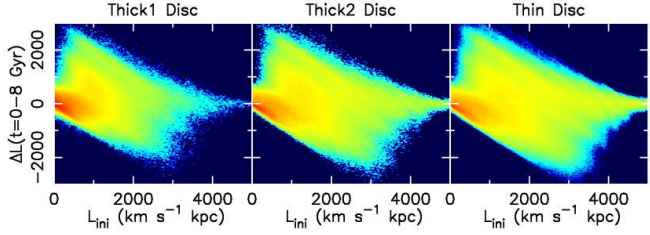}
    \caption{The change in angular momentum between $t=0$ and 8 Gyr as a function of the initial angular momentum for thick1 (left), thick2 (middle) and thin (right) disc components. The colour indicates the particle density normalised with the number of particles for each component (redder colour corresponds to higher density).}
    \label{fig:smapdL}
\end{figure}

 \begin{figure}
  \includegraphics[width=\hsize]{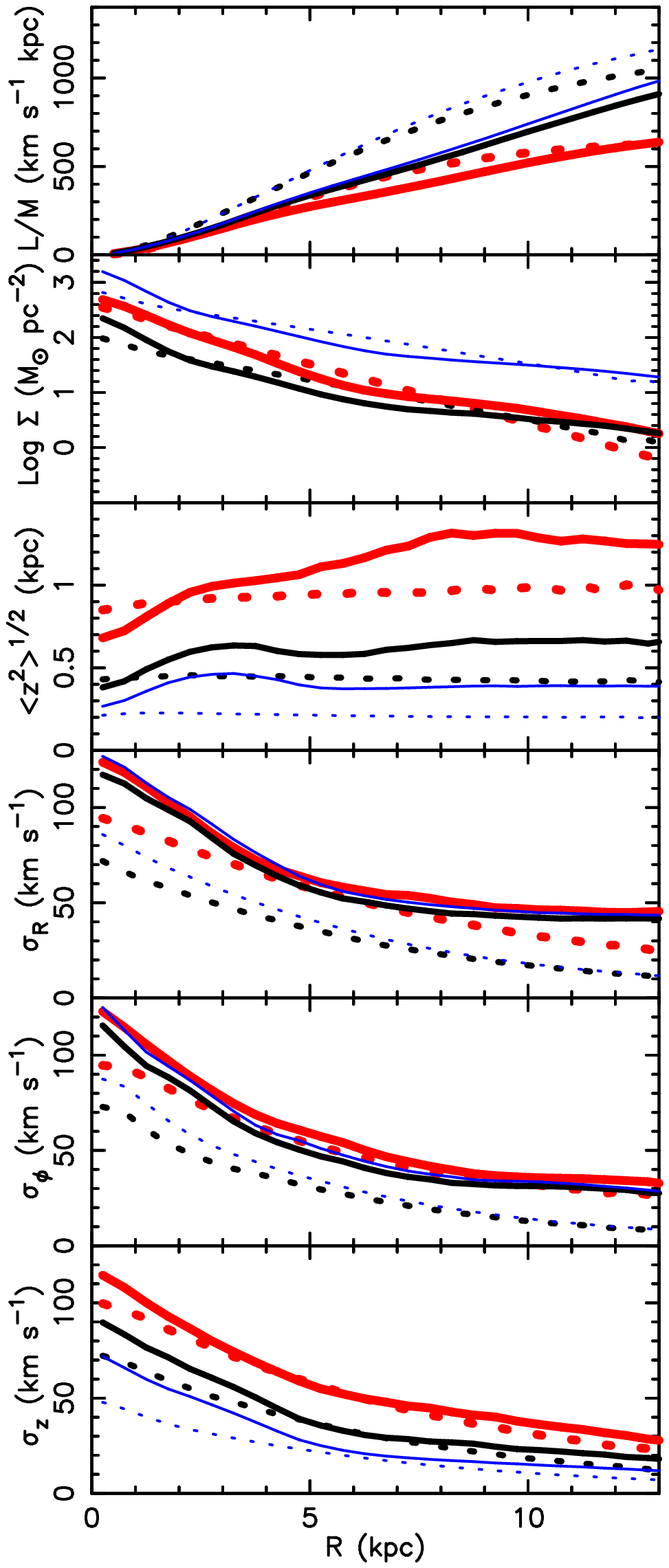}
    \caption{Radial profiles of cumulative specific angular momentum (top), surface mass density (2nd), root mean square height (3rd) and velocity dispersion of radial (4th), azimuthal (5th) and vertical (bottom) direction for thick1 (thicker red), thick2 (thick black) and thin (thin blue) disc components at $t=0$ (dashed lines) and $t=8$~Gyr (solid lines).}
    \label{fig:prof}
\end{figure}

\section{Results}
\label{sec:res}

Fig.~\ref{fig:snap} shows the evolution of the simulated galaxy. The system fully develops a central bar by $t=3.0$ Gyr, because of the dominated stellar mass compared to the dark matter mass in the central region. Fig.~\ref{fig:snap} shows that both thick disc components also develop a bar, although it is rounder than the thin disc one. This is consistent with what is shown in \citet{bt11,kggchb17}. Fig.~\ref{fig:smapdL} presents the angular momentum change between $t=0$ and $t=8$ Gyr, i.e. churning which includes both radial scattering and co-rotation radial migration. As expected, churning is relatively stronger for the thin disc than the thick discs. However, there are significant churning in both the thick1 and thick2 discs. Because the thick2 disc is kinematically colder than the thick1 disc, more significant churning can be seen in the thick2 disc than in the thick1 disc \citep[cf.][]{sss12}.

Fig.~\ref{fig:prof} shows the radial profile of the cumulative specific angular momentum, $L/M$, surface density, $\Sigma$, root mean square (RMS) of the height, $<\,z^2>^{1/2}$, and velocity dispersions of the radial, $\sigma_R$, azimuthal, $\sigma_{\phi}$, and vertical, $\sigma_z$, directions for the thin (blue), thick1 (black) and thick2 (red) disc components at $t=0$ (dotted lines) and 8 (solid lines) Gyr. Due to bar formation, all disc particles in the radial range plotted in Fig.~\ref{fig:prof} lost angular momentum so the surface density is more centrally concentrated at $t=8$~Gyr. There is a slight increase in surface density in the outer disk $R>8$~kpc especially for the thick1 disc. This is because thick1 disc was initially the most compact component, and more stars gained angular momentum compared to the other components as seen in the top panel. Hence, radial mixing driven by a larger disc component can result in radial expansion of the more compact disc component, though the effect is not significant \citep[see also][]{abs16}.

For all the components, the RMS height is smaller in the central regions, as a result of the deeper gravitational potential. In the outer region, $R>\sim3$ kpc, the RMS height becomes larger compared to the initial state. This is mainly due to the combination of heating, which can be seen as an increase in $\sigma_z$ in the bottom panel, and enhanced vertical oscillations of the stars that migrated outward. The latter mechanism is caused by higher initial vertical action for the outward migrators than the non-migrated ones, because of the deeper gravitational potential in the inner region \citep[e.g.][]{mfqdms12}. Interestingly, thick1 disc shows an obvious increase in $<\,z^2>^{1/2}$ with radius, i.e. flaring. The radial dependence of the scale-height for the thick disc is still under debate. Using APOGEE data, \citet{brsnhsb16} showed that the chemically-defined thick disc has a constant scale-height, independent of radii, as inferred from the  analysis of mono-abundance populations of the Galactic disc. On the other hand, \citet{mscmamdj17} argued that the mono-abundance populations are not same as the mono-age populations, and the old chemically-defined thick disc population can have a flared disc. This idea is supported by \citet{mbszcfgp17}, who analysed mono-age components of mono-abundance populations in the APOGEE data, and found that the old mono-age thick disc component shows flaring \citep[see also][]{rrfcrm14}. Although the structure of our disc components is beyond the scope of this study, it is interesting to note that, in our simulations, radial mixing leads to a flaring thick disc.

 The thin disc of our simulation is too hot compared to the velocity dispersion of stars in the solar neighbourhood. Our thin disc stars in $7.5<R<8.5$~kpc and $-0.5<z<0.5$~kpc have velocity dispersions of $(\sigma_R, \sigma_{\phi}, \sigma_{z})=(48.2, 36.5, 16.3)$~km~s$^{-1}$, while the solar neighbourhood thin disc stars defined as [$\alpha$/Fe]$<0.1$ measured in Gaia DR1 combined with APOGEE data shows $(37\pm2, 23\pm1, 18\pm1)$~km~s$^{-1}$ \citep{apkc16}. In this idealised numerical experiment, we do not include newly formed stars, and those would add kinematically colder stars to the thin disc component \citep[e.g.][]{jsrc84,abs16}. This simplification leads to a velocity dispersion of the thin disc that is too high at $t=8$~Gyr. In this paper, we focus on the thick disc, and the thin disc component is merely used as a driver of radial mixing. We do not analyse the properties of the thin disc component, and therefore, this discrepancy is unlikely to be an issue for our qualitative study of the effect of radial mixing on the metallicity distribution of the thick disc. 
 
   The relatively high velocity dispersion mentioned above implies that our N-body simulation likely overestimates radial mixing by radial scattering mainly due to the bar formation, compared to what is expected to happen in the Milky Way. On the other hand, a high velocity dispersion weakens the spiral arms and reduces co-rotation radial migration, which underestimates radial mixing. Because we do not know what is the degree of radial mixing in the Milky Way, it is difficult to assess whether or not our simulation has an appropriate level of radial mixing for the thick disc compared to the Milky  Way. However, as seen in Fig.~\ref{fig:smapdL}, both thick disc components experience significant radial mixing. Hence, our numerical experiment is suitable to provide a qualitative evaluation of how radial mixing affects the radial and vertical metallicity gradient, but we will avoid a quantitative discussion.
   


\begin{figure}
  \includegraphics[width=\hsize]{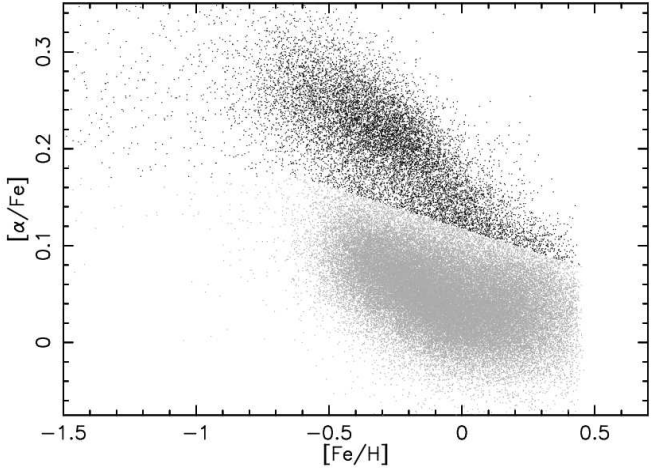}
    \caption{[$\alpha$/Fe] as a function of [Fe/H] for APOGEE data from \citet{hbhnb15}. Black (grey) dots indicate the selected thick (thin) disc stars.}
    \label{fig:H15alfe_c}
\end{figure}

 Fig.~\ref{fig:H15alfe_c} shows the [$\alpha$/Fe]  and [Fe/H] distribution of APOGEE stars in Sloan Digital Sky Survey (SDSS) data release 12 from \citet{hbhnb15}. Although the data show two distinct sequences for the [$\alpha$/Fe] and [Fe/H] relation, there is no obvious definition for chemically identified thick and thin disc stars. In this paper, we define thick disc stars to be ${\rm [\alpha/Fe]}>0.17$ at [Fe/H]$<-0.6$ and ${\rm [\alpha/Fe]}>-0.0875 (\text{[Fe/H]}-0.2)+0.1$ at [Fe/H]$>-0.6$, which shows a reasonable division of the sequences in Fig.~\ref{fig:H15alfe_c}. We consider that these chemically-defined thick disc stars represent the old thick disc population, and we refer to them as the APOGEE thick disc stars. We explore what radial and vertical metallicity gradients are allowed in the thick disc when is born, to explain the MDF of the APOGEE thick disc stars at different radii and vertical heights.
 

\begin{table*}
 \caption{Results of the MCMC fitting}
 \label{tab:MCMC-res}
 \begin{tabular}{lccccccccc}
  \hline
 & \multicolumn{4}{c}{Thick1+Thick2 or Thick1 only} & & \multicolumn{4}{c}{Thick2} \\
  \cline{2-5} \cline{7-10} 
Case & [Fe/H]$_{0}$ & (d[Fe/H]/dR)$_0$ & (d[Fe/H]/dz)$_0$ & $\sigma_{{\rm [Fe/H]},0}$ & &
   [Fe/H]$_{0}$ & (d[Fe/H]/dR)$_0$ & (d[Fe/H]/dz)$_0$ & $\sigma_{{\rm [Fe/H]},0}$  \\
& dex & dex kpc$^{-1}$ & dex kpc$^{-1}$ & dex & & 
    dex & dex kpc$^{-1}$ & dex kpc$^{-1}$ & dex \\
 \hline
C1  & $-0.289\pm 0.016$ & $0.023\pm 0.002$ & $-0.242\pm 0.008$ & $0.192\pm 0.003$ & & 
  $-$ & $-$ & $-$ & $-$ \\
C2 &  $-0.271\pm 0.022$ & $0.012\pm 0.004$ & $-0.206\pm 0.012$ & $0.151\pm 0.006$ &&
   $0.451\pm0.033$ & $-0.048\pm 0.005$ & $-0.726\pm 0.040$ & $0.131\pm 0.014$ \\
  \hline
 \end{tabular}
\end{table*}

\begin{table}
 \caption{Radial and Vertical Metallicity Gradients.}
 \label{tab:metgrad}
 \begin{tabular}{lcc}
  \hline
Case &  d[Fe/H]/dR$^a$ & d[Fe/H]/dz$^b$ \\
& dex kpc$^{-1}$ & dex kpc$^{-1}$ \\
 \hline
 APOGEE & $0.0005\pm0.0014$ &  $-0.1760\pm0.0068$ \\
C1($t=8$) & $0.0144\pm0.0001$ & $-0.1173\pm0.0015$ \\
C2($t=8$) & $0.0020\pm0.0001$ & $-0.1477\pm0.0015$ \\
C2($t=0$) & $-0.0066\pm0.0006$ & $-0.3537\pm0.0014$ \\
C2 thick1($t=8$) & $0.0072\pm0.0001$ & $-0.0757\pm0.0015$ \\
C2 thick2($t=8$) & $-0.0204\pm0.0002$ & $-0.1717\pm0.0037$ \\
  \hline
 \end{tabular}
 \\
 $^a$ Radial metallicity gradient at $3<R<13$~kpc and $|z|<2.0$~kpc.\\
 $^b$ Vertical metallicity gradient at $7<R<9$~kpc and $|z|<2.0$~kpc.
\end{table}

\subsection{Single metallicity distribution case}
\label{sec:onedisc}

\begin{figure*}
  \includegraphics[width=\hsize]{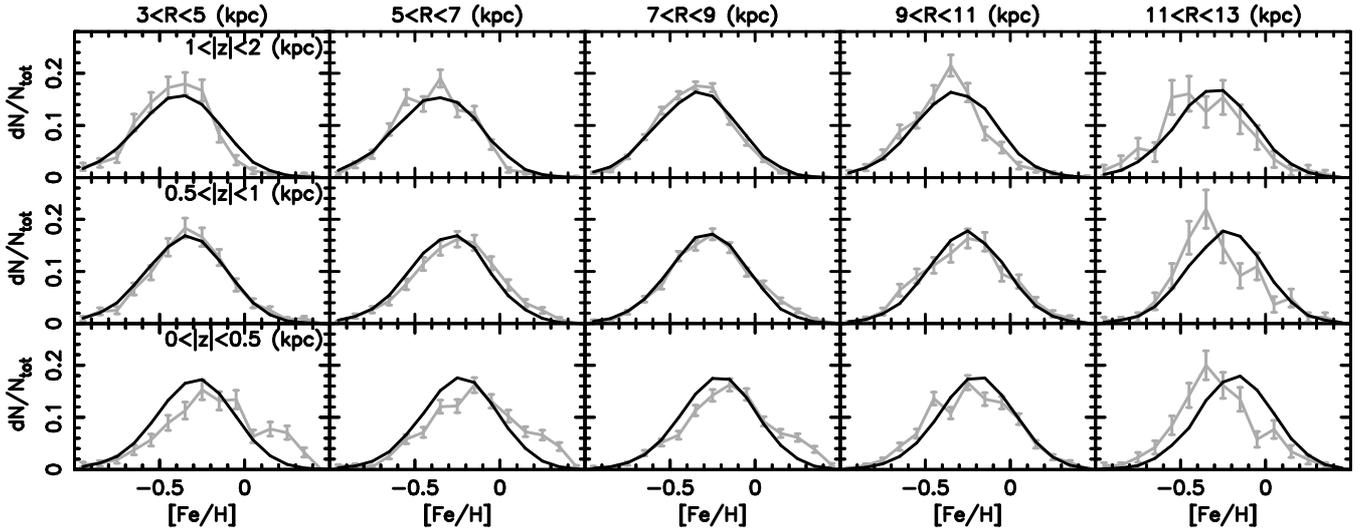}
    \caption{Metallicity distribution function (MDF) of the APOGEE thick disc stars (grey lines with error bars) and the best fit model of Case C1 (black solid line) at different radial and heights. The top, middle and bottom panel correspond to MDFs in $0<|z|<0.5$, $0.5<|z|<1$ and $1<|z|<2$~kpc volumes, respectively. Left, 2nd-left, middle, 2nd-right and right panels show MDFs in the radial region of $3<R<5$, $5<R<7$, $7<R<9$, $9<R<11$ and $11<R<13$~kpc, respectively.}
    \label{fig:C1-mdf}
\end{figure*}

 First, we assume both the thick1 and thick2 discs to have the same initial metallicity distribution. We call this Case C1. Grey lines and error bars in Fig.~\ref{fig:C1-mdf} represent the MDF for the APOGEE thick disc stars at different radial and height regions. We divide the sample of stars within $3<R<13$~kpc with a 2~kpc bin size, and 3 different height bins of $0<|z|<0.5$, $0.5<|z|<1$ and $1<|z|<2$~kpc. The error bars correspond to Poissonian statistical errors. We run MCMC with 4 parameters, the initial $R=0$ and $|z|=0$ metallicity, [Fe/H]$_0$, radial metallicity gradient, (d[Fe/H]/dR)$_0$, vertical metallicity gradient, (d[Fe/H]/dz)$_0$, and metallicity dispersion, $\sigma_{\rm [Fe/H],0}$, at $t=0$, to explore what parameter set would best match with the APOGEE thick disc MDFs. For each MCMC chain, we assign [Fe/H] to particles using the position at $t=0$ to follow the parameters of [Fe/H]$_0$, (d[Fe/H]/dR)$_0$, (d[Fe/H]/dz)$_0$ and $\sigma_{\rm [Fe/H],0}$. We also add Gaussian random errors to [Fe/H] of $0.05$~dex to take into account the typical observational errors for APOGEE data \citep{hbhnb15}. Then, using their position at $t=8$~Gyr, we analyse the MDF in the same volumes as for the APOGEE thick disc data in Fig.~\ref{fig:C1-mdf}, and calculate the following log likelihood function.
\begin{eqnarray}
\ln L &= & -\frac{1}{2}\sum_i \left[ \frac{(dN({\rm [Fe/H]}_{i,{\rm model}})-dN({\rm [Fe/H]}_{i,{\rm obs}}))^2}{\sigma_{\rm [Fe/H]_{{\it i},obs}}^2} \right. \nonumber \\
& & \left. + \ln(2 \pi \sigma_{\rm [Fe/H]_{{\it i},obs}}) \right],
\label{eq:like}
\end{eqnarray}
where $dN({\rm [Fe/H]_{\it i}})$ are the normalised number of APOGEE thick disc stars or model particles which are in the $i$-th [Fe/H] bin, and $i$ indicates each [Fe/H]$_i$ bin in the MDFs for all different volumes described above, except at the location where $dN({\rm [Fe/H]}_{i,{\rm obs}})=0$. We define the observational errors, $\sigma_{\rm [Fe/H]_{{\it i},obs}}$, of the $dN({\rm [Fe/H]_{\it i}})$ as Poisson errors from the number of stars in each bin. We use flat priors in the ranges of $-2.0<{\rm [Fe/H]}_0<2.0$~dex, $-0.5<{\rm (d[Fe/H]/dR)_0}<0.5$~dex~kpc$^{-1}$, $-1.0<{\rm (d[Fe/H]/dz)_0}<1.0$~dex~kpc$^{-1}$ and $0.0<\sigma_{\rm [Fe/H],0}<1.0$~dex. We use {\tt emcee} \citep{fmhlg13} for the MCMC sampler, and use 128 walkers and 2,000 chains. We call this  assigning [Fe/H] to particles with "MCMC chemical painting" or "MCMC painting". Note that chemically painting N-body particles has been used in \citet{mcm13,mcm14,kpa13,kpa15a} where a chemical evolution model is used to assign metallicities to the particles. Our method is simpler, but allows us to fit the observational data with MCMC in a reasonable computational time. Hence, these different methods are complementary to each other.
 
The best fit parameters found with the MCMC chemical painting analysis are shown in Table~\ref{tab:MCMC-res} and the MDFs of the best model are shown in Fig.~\ref{fig:C1-mdf} (black lines). Our MCMC painting analysis for Case~C1 suggests that a positive radial metallicity gradient of (d[Fe/H]/dR)$_0=0.023\pm0.002$~dex~kpc$^{-1}$ for the initial disc is preferred to reproduce the observed MDF. A positive metallicity gradient may come as a surprise, because this is rare in disc galaxies. However, we find that this is required to reproduce the flat radial metallicity gradient and the negative vertical metallicity gradient observed in the thick disc of the Milky Way at present.

We measured the radial metallicity gradient of the APOGEE thick disc stars in the region $3<R<13$~kpc and the vertical height within $|z|<2$~kpc, and found d[Fe/H]/dR$=0.0005\pm0.0014$~dex~kpc$^{-1}$ as shown in Table~\ref{tab:metgrad}. We also measured the vertical metallicity gradient in the region of $7<R<9$~kpc, and found d[Fe/H]/dz$=-0.1760\pm0.0068$~dex~kpc$^{-1}$  at $|z|<2$~kpc. As mentioned in Section~\ref{sec:intro}, the thick disc stars have a flat radial metallicity gradient and a clear negative vertical metallicity gradient. We also measured the radial metallicity gradients in the same regions with the same way for our best fit model at $t=8$~Gyr (Tab.~\ref{tab:metgrad}). We found d[Fe/H]/dR$=0.0144\pm0.0001$~dex~kpc$^{-1}$, which is reasonably similar to  the flat radial metallicity gradient found from the APOGEE thick disc stars\footnote{ Interestingly, there is a hint of positive radial metallicity gradient of the old thick disc discussed in Fig.~18 of \citet{csacrmf11}.}. Hence, radial mixing induced by the bar and spiral arms seems to have effectively flattened the radial metallicity gradient from the initially assumed positive metallicity gradient. An important mechanism is that the stars brought from the inner (outer) region due to the radial mixing tend to populate the high (low) vertical height region, as shown in \citet{kggchb17}. Hence, if there is initially a negative radial metallicity gradient, radial mixing will make the vertical metallicity gradient positive, which is inconsistent with the clear negative vertical metallicity gradient seen in the APOGEE data. This is why our MCMC result arrived at a positive radial metallicity gradient for the initial disc; to reproduce the negative vertical metallicity gradient after radial mixing brought stars from the inner disc and populated with them the regions far from the plane. This result suggests that the negative vertical metallicity gradient observed in the thick disc of the Milky Way provides a strong constraint on the overall radial metallicity gradient of the thick disc progenitor and calls for a positive radial metallicity gradient. 

 Fig.~\ref{fig:C1-mdf} shows the best model MDF misses higher metallicity stars at $|z|<0.5$~kpc, which explains the shallower vertical metallicity gradient of the best model of Case~C1 compared to the APOGEE thick disc data, as seen in Table~\ref{tab:metgrad}. Still, this is the best model found by our MCMC run with the one component model. This model is informative to understand the general requirement to explain the observed MDF of the APOGEE thick disc data. However, this inconsistency indicates that we need more components to explain the observed MDF. Hence, in the next section, we consider a two component model.

\begin{figure*}
  \includegraphics[width=\hsize]{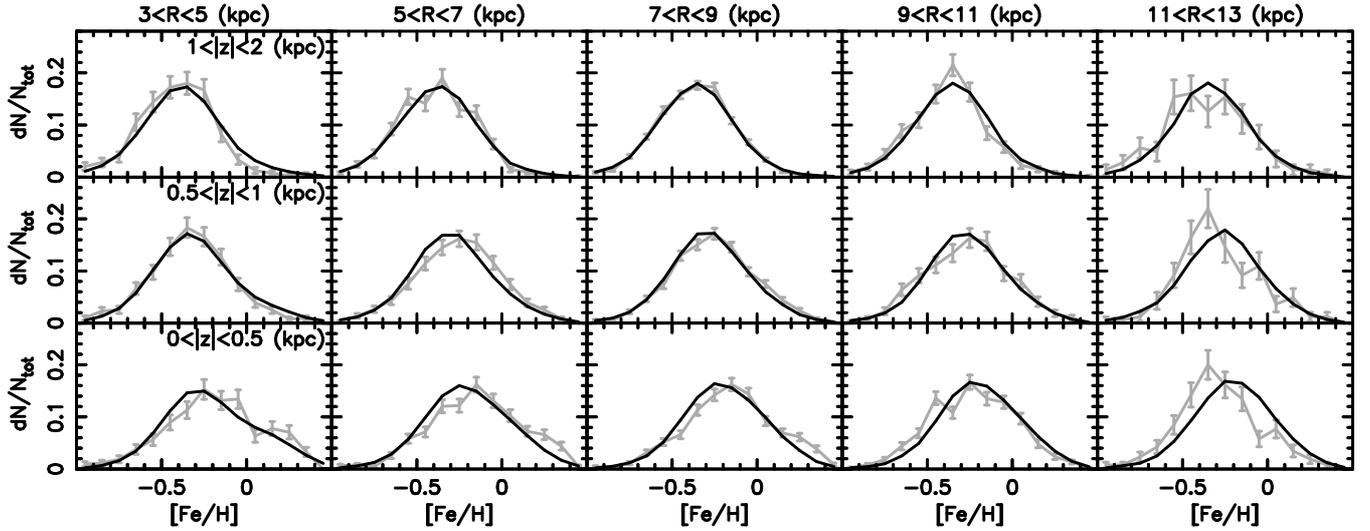}
    \caption{Same as Fig.~\ref{fig:C1-mdf} but for Case C2.}
\label{fig:C2-mdf}
\end{figure*}

\begin{figure}
  \includegraphics[width=\hsize]{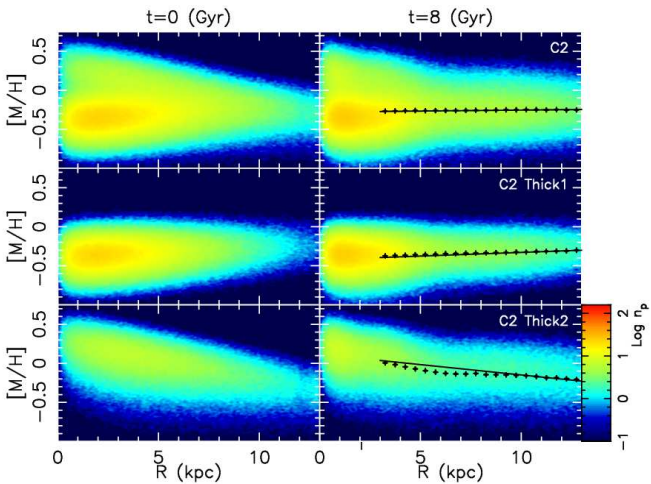}
    \caption{Radial metallicity distribution of the whole disc (top) and thick1 (middle) and thick2 (bottom) disc components for Case C2 at $t=0$ (left) and $t=8$ (right) Gyr. The points with error bars in the right panels show the mean metallicity of each radial bin which we used for a linear regression (solid line) summarised in Table.~\ref{tab:metgrad}. The error bars describe the uncertainty in the mean, i.e. $\sigma/\sqrt{N}$, where $\sigma$ is dispersion and $N$ is number of particle in each bin, which are very small.}
\label{fig:C2-rmet}
\end{figure}

\subsection{Double metallicity distribution case}
\label{sec:twodisc}

 We now consider the thick1 and thick2 discs separately, i.e. we allow different radial and vertical metallicity gradients to be assigned for these two discs. We run MCMC with 8 parameters, i.e. two different sets of [Fe/H]$_0$, (d[Fe/H]/dR)$_0$, (d[Fe/H]/dz)$_0$ and $\sigma_{\rm [Fe/H],0}$ for thick1 and thick2 discs at $t=0$, and analyse the log likelihood in the same way as equation~(\ref{eq:like}) in previous section. We set the same flat priors as mentioned in Section~\ref{sec:onedisc}. We call this Case~C2.
 
The best fit parameters for Case~C2 are summarised in Table~\ref{tab:MCMC-res}, and the MDFs of the best fit model are shown in Fig.~\ref{fig:C2-mdf}. Fig.~\ref{fig:C2-mdf} shows that the two component disc case reproduces better the APOGEE thick disc data than Case~C1. The higher metallicity tails of the MDFs at $|z|<0.5$~kpc and $R<9$~kpc are described better. Overestimates of high metallicity stars at $1<|z|<2$~kpc and $R>11$~kpc in Case~C1 are mitigated in Case~C2. 

There are some regions where our best fit model of Case~C2 still struggles to reproduce the observed MDF. However, it is not our aim to reproduce the observed MDF perfectly. We are using only one N-body simulation result with one evolution history. The aim of this paper is to qualitatively discuss the effect of radial mixing on the thick disc metallicity distribution, and not to find the best evolution history of the Galactic disc. Therefore, we consider that we have a reasonable success in qualitatively reproducing the APOGEE data, and discuss what we can learn from the fact that two component case is better than one component case. 

Table~\ref{tab:MCMC-res} shows that to reproduce the APOGEE thick disc MDF, the thick1 disc should have a positive radial metallicity gradient of (d[Fe/H]/dR)$_0=0.012\pm 0.004$~dex~kpc$^{-1}$, like Case~C1, while the thick2 disc should have a negative radial metallicity gradient of (d[Fe/H]/dR)$_0=-0.048\pm 0.005$~dex~kpc$^{-1}$. Also, the slope required for the thick1 disc is shallower than the slope required for Case~C1. Note that the thick1 disc is more massive than the thick2 disc, and the thick1 disc is thicker in scale height and smaller in scale radius. Our best fit MCMC result shows that the thick2 disc should be more metal rich than the thick1 disc. Hence, if we combine the thick1 and thick2 discs, the inner, higher regions are dominated by the metal poor thick1 disc, and the outer region and lower height region is more dominated by the metal rich thick2 disc. As a result, this combined disc provides a similar metallicity distribution to a single positive radial metallicity gradient in Case~C1. We measured the overall radial metallicity gradient at $t=0$ for the best model of Case~C2 in the same way as Section~\ref{sec:onedisc}, and found d[Fe/H]/dR$=-0.0066\pm0.0006$, which is almost flat or slightly negative (labeled "C2 (t=0)" in Table~\ref{tab:metgrad}). However, note that we measured the metallicity gradients in Table~\ref{tab:metgrad} by a linear regression of the mean [Fe/H] of all the particles of both thick1 and thick2 discs at radial and vertical bins. In other words, this does not take into account the density variations with radii. Hence, this result likely underestimates the contribution of the metal poor particles in the inner region from the thick1 disc. Fig.~\ref{fig:C2-rmet} shows the radial metallicity distribution for the whole disc and thick1 and thick2 disc components at $t=0$ and 8 Gyr. In fact, metal poor particles are dominant in the inner region, but they are not contributing to our linear regression, because our linear regression is done outside of this region, as seen in the top right panel of Fig.~\ref{fig:C2-rmet}. We therefore consider that, overall, the best model in Case C2 has more metal poor stars in the inner region and more metal rich stars in the outer region. As discussed in Section~\ref{sec:onedisc}, this is required to reproduce the negative vertical metallicity gradient in the APOGEE data. 

 We conclude that the thick disc of the Milky Way can be explained with two or multiple generations of the discs. Considering the age-metallicity relation (older thick disc stars having lower [Fe/H]) observed in the thick disc \citep{hdmlkg13,bfo14,brsfa14}, we can suggest a scenario where a thicker and smaller thick1-like disc formed first, and then later a more metal rich thinner and larger thick2-like disc was built up. Alternatively and more naturally, the thick disc was built up by gradually increasing in radial size and metallicity, and decreasing in thickness, i.e. an inside-out and upside-down fashion \citep{bkmg06,jbkw12}. In the latter scenario, the earlier generations of the thick disc are more metal poor and more compact, which leads to an overall positive radial metallicity gradient, despite individual generations of the disc can have had a flatter or even negative radial metallicity gradient. This is first shown in Fig.~10 of \citet{mcm14} and also discussed in \citet{rspm17}. These results mean that the inside-out formation can allow the thick disc star-forming region to have a variety of radial metallicity gradients. This helps to explain the range of radial metallicity gradients observed in high redshift disc galaxies, as mentioned in Section~\ref{sec:intro}. On the other hand, radial mixing limits the range of radial metallicity gradient possible for the thick disc progenitor, since the current thick disc has a clear negative vertical metallicity gradient. 

It is worth noting that the preferred solution for the thick2 disc is an initial steep vertical metallicity gradient of d[Fe/H]/dz$=-0.726\pm0.040$. This is required to reproduce the negative vertical metallicity gradient in APOGEE data, because the negative radial metallicity gradient and radial mixing make the vertical metallicity gradient more positive. It is difficult to imagine the star forming gas disc having such a steep vertical metallicity gradient, especially if turbulent mixing is strong in high redshift galaxies \citep{ccymk12,gkc14,ptgf15}. If we assume that metal mixing is effective and there should be no vertical metallicity gradient in the star forming disc at a fixed time, the steep vertical metallicity gradient indicates that the coeval population of the thick disc progenitor forming at different epochs gradually became thinner with time, and later populations are more metal rich. Then, the overall negative vertical metallicity gradient can be established by the thick disc built up in an upside-down fashion, with younger populations having higher [Fe/H]. Hence, to describe the thick disc formation, we need to consider multiple-generations of discs with different radial sizes and thicknesses and different metallicities or smooth growth of the disc, rather than only a two component disc.

\subsection{[Fe/H] and the Rotation Velocity Relationship}

\begin{figure*}
  \includegraphics[width=\hsize]{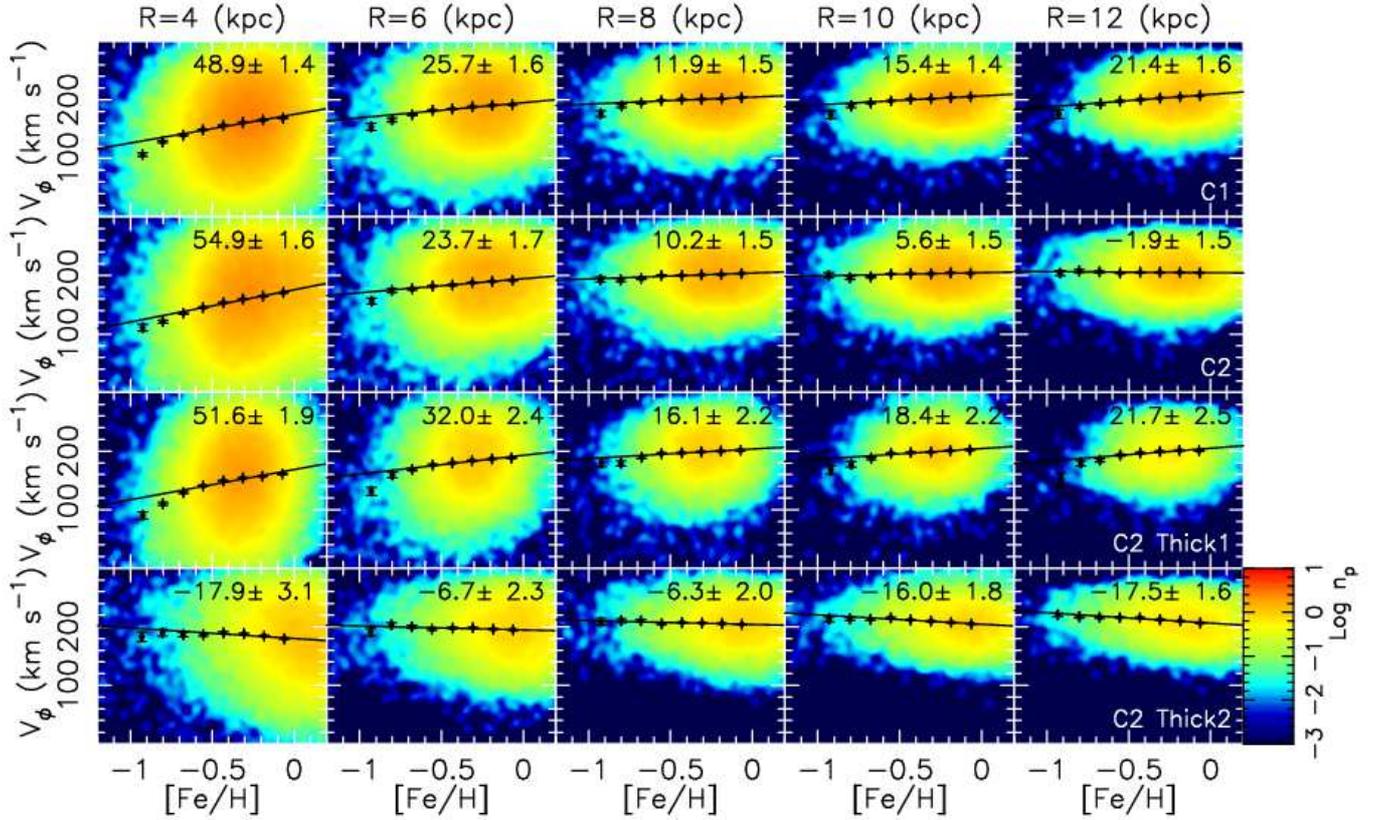}
    \caption{Rotation velocity, $V_{\phi}$ as a function of [Fe/H] for the thick disc particles for the best model at the radius of of $R=4$ (left), 6 (2nd left), 8 (middle), 10 (2nd right) and 12 (right) kpc for Case C1 (top), Case C2 (2nd) and thick1 (3rd) and thick2 discs in Case C2 (bottom). }
\label{fig:smapvphiZ}
\end{figure*}

\begin{figure}
  \includegraphics[width=\hsize]{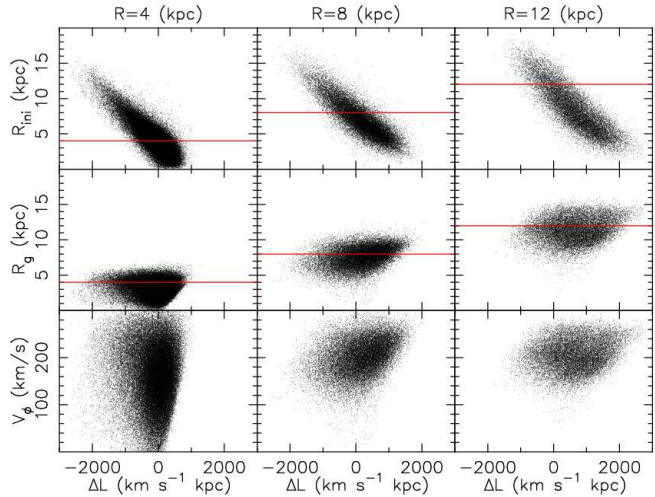}
    \caption{Radius at $t=0$, $R_{\rm ini}$ (top), guiding radius, $R_{\rm g}$ (middle) and the rotation velocity, $V_{\phi}$ (bottom) as a function of the change in angular momentum between $t=0$ and 8 Gyr for the particles at $R=4$ (left), 8 (middle) and 12 (right)~kpc for Case~C1. The horizontal red lines in the top and middle panels show the current radius, i.e. $R=4, 8$ and 12~kpc for the left, middle and right panels, respectively}
\label{fig:dLRV}
\end{figure}

The thick disc stars defined by their [$\alpha$/Fe] abundance have a distinctive feature in the [Fe/H]-$V_{\phi}$ relation, compared to the chemically-defined thin disc in the solar neighbourhood stars. Observational studies demonstrate that thick disc stars have a positive slope of d$V_{\phi}$/d[Fe/H], while the thin disc stars show a negative slope \citep[e.g.][]{slrfs10,lbaij11,afshsp13,rbdlkhhg14,apkc16}. There are several theoretical possibilities to explain this trend \citep[e.g.][]{lrdiqw11,clsmmrfs12,apkc16,rspm17}. In this section, we discuss the [Fe/H]-$V_{\phi}$ relation for the thick disc only, given it is the focus of this paper. Note that the negative gradient d$V_{\phi}$/d[Fe/H] for the thin disc can be easily explained by the observed radial metallicity gradient and stellar epicycle motion, i.e. blurring \citep{vcdna14,apkc16}.

Fig.~\ref{fig:smapvphiZ} shows the rotation velocity, $V_{\phi}$, as a function of [Fe/H] for the thick disc particles for the best model for Cases~C1 and C2 and thick1 and thick2 discs in Case~C2. As discussed in \citet{clsmmrfs12,apkc16,rspm17}, a positive slope of d$V_{\phi}$/d[Fe/H] can be explained with an overall positive radial metallicity gradient, which explains the clear positive slope in Case~C1 and thick1 disc in Case~C2. In these cases, the stars whose guiding radius is smaller (larger) are more metal poor (rich), and tend to be rotating slower (faster), because they tend to be near apocentre (pericentre) phase. This is due to the blurring effect, not due to churning, i.e. the change in angular momentum. The steeper slope in the inner region is due to the increase in the velocity dispersion in the inner disc (see Fig.~\ref{fig:prof}), which leads to a stronger change in $V_{\phi}$ at a given radius. 

It is interesting that in Case~C1 and thick1 disc in Case~C2 the slope of d$V_{\phi}$/d[Fe/H] is minimum at $R=8$~kpc, and becomes steeper at $R>8$ kpc. We found that this is because around $R=8$~kpc churning is most effective. The top panels of Fig.~\ref{fig:dLRV} show the initial radii, $R_{\rm ini}$, of particles at $R=4$, 8 and 12~kpc at $t=8$ Gyr. In the inner (outer) region, i.e. at $R=4$ (12)~kpc, more particles came from the outer (inner) disc, i.e. $R_{\rm ini}$ is larger (smaller) than the current radius, because of loss (gain) of angular momentum. On the other hand, at $R=8$~kpc the particles came from both inner and outer region. We calculated the guiding radius, $R_{\rm g}$, at $t=8$~Gyr, by comparing angular momentum of the particles and the circular velocity at the disc plane, which are shown in the middle panels of Fig.~\ref{fig:dLRV}. It is interesting to see that the guiding radius tends to be smaller (higher) than the current radius for the particles which lost (gained) their angular momentum. For example, for the particles at $R=8$~kpc, those with negative (positive) $\Delta L$ tend to have smaller (larger) $R_{\rm g}$ than the red line of $R=8$~kpc in the middle panel of Fig.~\ref{fig:dLRV}. As a result, the particles that lost (gained) angular momentum tend to be detected at apocentre (pericentre) phase, and therefore are rotating more slowly (faster). The particles from larger (smaller) $R_{\rm ini}$ are more metal rich (poor), because of the initial positive radial metallicity gradient in Case C1. Consequently, the particles which changed their guiding centre from the outer (inner) region to the inner (outer) region contributes more to the metal rich (poor) and slowly (fast) rotating region in the [Fe/H]-$V_{\phi}$ relation, flattening a positive slope of d$V_{\phi}/$d[Fe/H]. This result implies that churning can flatten a positive d$V_{\phi}$/d[Fe/H] slope. We also confirmed that if we picked up particles with small $|\Delta L|$, the d$V_{\phi}$/d[Fe/H] slope is more positive in Case~C1. Because at $R=8$~kpc there are particles that came from both sides of the disc, d$V_{\phi}$/d[Fe/H] becomes more flatten by churning. Some indication of this kind of trend is observed in \citet{kwcmas17}.

It is interesting that the thick2 disc shows negative d$V_{\phi}$/d[Fe/H] slopes at all radii. It is basically the same mechanism as the thin disc case as discussed above, i.e. blurring and negative radial metallicity gradient drives a negative d$V_{\phi}$/d[Fe/H] slope. Because the outer region is dominated by the thick2 disc in Case~C2, the slope of d$V_{\phi}/$d[Fe/H] becomes shallower, or even negative, in the outer disc (2nd top panels in Fig.~\ref{fig:smapvphiZ}). We expect that if the thick disc was built up in an inside-out fashion and the star forming disc at different epochs had a negative radial metallicity gradient, an mono-age thick disc population at $z=0$ could have a negative d$V_{\phi}$/d[Fe/H] like the thick2 disc. Then, the overall d$V_{\phi}/$d[Fe/H] as a function of radius for the thick disc depend on the inside-out formation history and the radial metallicity gradient for each mono-age population.

At $R=8$~kpc, our models show too shallow slope of d$V_{\phi}$/d[Fe/H], compared to the observed one for the Galactic thick disc, e.g.\ one of the shallowest slope of d$V_{\phi}$/d[Fe/H]$=23\pm10$~km~s$^{-1}$ is reported in \citep{apkc16}. If the thick disc consists of many generations of discs with different thickness, scale-length and [Fe/H], we can change this slope by adjusting it to reproduce the observed slopes, as \citet{rspm17} demonstrated. However, such work is beyond the scope of this paper. Our qualitative study demonstrates that an overall positive d$V_{\phi}$/d[Fe/H] slope can be driven by an overall positive radial metallicity gradient, and radial mixing of churning does flatten this slope. Hence, the measurement of d$V_{\phi}$/d[Fe/H] as a function of radius for different mono-age populations provides strong constraints on the formation history of the thick disc. 

\section{Summary and Discussion}
\label{sec:sum}

Combining an idealised N-body model of a disc galaxy similar in size to the Milky Way with a new MCMC chemical painting technique, we explore what metallicity distribution for the thick disc progenitor is needed, to explain the current thick disc metallicity distribution observed by the APOGEE survey. The main assumptions in our study are that thick disc formation was completed a long time (8~Gyr) ago and the thick disc stars experienced effective radial mixing for a long time since then. In addition, we assumed that the vertical scale height of the overall thick disc component is constant at different radii. Our key findings are the followings. 

(i) The current flat radial metallicity gradient and negative vertical metallicity gradient of the thick disc stars observed in the APOGEE data provide strong constraints on the radial metallicity gradient for the thick disc progenitor; it should be flat or even positive. A negative radial metallicity gradient is not likely. This is because radial mixing with a negative radial metallicity gradient leads to a positive vertical metallicity gradient as demonstrated in \citet{kggchb17}, which is in conflict with observations. 

(ii) Two or more components for the thick disc can explain better the MDF of the thick disc stars. Our result supports a scenario in which the thick disc formed in an inside-out and upside-down fashion, and an older, thicker and smaller disc is more metal poor. In this case, each generation of thick disc could have an initial negative radial metallicity gradient. Then, inside-out formation and age-metallicity relation produced an overall flat or positive metallicity gradient when their formation was completed. 

(iii) Hence, the inside-out formation can allow the thick disc star-forming region to have a variety of (positive, flat or negative) radial metallicity gradients. This helps to explain the variety of radial metallicity gradients observed at high redshift disc galaxies. On the other hand, radial mixing rather limits the range of allowed radial metallicity gradient as discussed in (i). 

(iv) The overall positive radial metallicity gradient, built-up by inside-out formation and the age-metallicity relation, helps to explain a positive d$V_{\phi}$/d[Fe/H] slope observed for the thick disc stars, as suggested by \citet{rspm17}. In this case, radial mixing flattens the positive d$V_{\phi}$/d[Fe/H] slope, especially around the solar radius. 

To our knowledge, (i) and (iii) are new finding, although (iii) can be inferred from the results of \citet{rspm17}. (ii) is consistent with the scenario discussed in the semi-analytic chemo-dynamical model of \citet{rspm17}  and in the cosmologically motivated chemo-dynamical model of \citet{mcm14}. The first part of (iv) is an independent confirmation of the scenario suggested by  \citet{rspm17}, but the radial dependence of the  d$V_{\phi}$/d[Fe/H] slope is discussed here for the first time. These results indicate that if the thick disc formed in an inside-out and upside-down fashion with a clear age-metallicity relation, different generations of the thick disc progenitor should have associated different chemo-dynamical structures, i.e. different sizes and radial metallicity gradients and different d$V_{\phi}$/d[Fe/H] slopes at different radii. If we could measure the age of the old stars accurately enough, we would expect that the older thick disc would be smaller and more metal poor, and each generation of the thick disc would exhibit a more negative radial metallicity gradient than the overall gradient of the thick disc. Also, d$V_{\phi}$/d[Fe/H] for a mono-age thick disc population would be flatter or even negative (if their radial metallicity gradient is also negative). 

 Unfortunately, the thick disc formation time-scale is relatively short \citep[a few Gyr, e.g.][]{hdmlkg13}. It may be challenging to divide the sample of thick disc stars into age bins fine enough to test this scenario. Hence, this would be a key challenge for future work on Galactic archaeology. The European Space Agency (ESA)'s \textit{Gaia} mission will soon provide us accurate distance and photometric information supplemented with metallicity measurements from high-resolution spectroscopic surveys. The data will bring age estimates for turn-off stars from well-calibrated isochrones. In addition, for both dwarf and giants, asteroseismology will provide age estimates from the high-quality light-curve data from the on-going K2 mission with \textit{Kepler} \citep[e.g.][]{shsjl15} and up-coming \textit{TESS} and \textit{Plato} missions \citep[e.g.][]{mcmdfgjk17}. Our study demonstrates that fitting such observational data with N-body simulations with MCMC painting would be a valuable tool to understand the building history of the Galactic disc. In this paper, we only used one simplified simulation model without any growth of the disc mass. Encouraged by the success of this study, we will further develop N-body disc models including the growth of the disc, as in \citet{abs16}, and run many different models with various formation scenarios. Combining N-body simulation models and MCMC painting against future observational data must help us to decipher the formation and evolution of the Milky Way disc.

 \section*{Acknowledgments}
We thank an anonymous referee for their constructive comments and helpful suggestions which have improved the manuscript. DK also thanks Ivan Minchev and Ralph Sch\"onrich for providing insightful comments.
DK and IC acknowledge the support of the UK's Science \& Technology Facilities Council (STFC Grants ST/K000977/1 and ST/N000811/1). CAP is thankful to the Spanish MINECO for funding through grant AYA2014-56359-P. LC gratefully acknowledges support from the Australian Research Council (grants DP150100250, FT160100402). RJJG acknowledges support by the DFG Research Centre SFB-881 'The Milky Way System', through project A1. JH is supported by a Dunlap Fellowship at the Dunlap Institute for Astronomy \& Astrophysics, funded through an endowment established by the Dunlap family and the University of Toronto. This work used the UCL facility Grace and the DiRAC Data Analytic system at the University of Cambridge, operated by the University of Cambridge High Performance Computing Service on behalf of the STFC DiRAC HPC Facility (www.dirac.ac.uk). This equipment was funded by BIS National E-infrastructure capital grant (ST/K001590/1), STFC capital grants ST/H008861/1 and ST/H00887X/1, and STFC DiRAC Operations grant ST/K00333X/1. DiRAC is part of the National E-Infrastructure.

Funding for SDSS-III has been provided by the Alfred P. Sloan Foundation, the Participating Institutions, the National Science Foundation, and the U.S. Department of Energy Office of Science. The SDSS-III web site is http://www.sdss3.org/.

SDSS-III is managed by the Astrophysical Research Consortium for the Participating Institutions of the SDSS-III Collaboration including the University of Arizona, the Brazilian Participation Group, Brookhaven National Laboratory, Carnegie Mellon University, University of Florida, the French Participation Group, the German Participation Group, Harvard University, the Instituto de Astrofisica de Canarias, the Michigan State/Notre Dame/JINA Participation Group, Johns Hopkins University, Lawrence Berkeley National Laboratory, Max Planck Institute for Astrophysics, Max Planck Institute for Extraterrestrial Physics, New Mexico State University, New York University, Ohio State University, Pennsylvania State University, University of Portsmouth, Princeton University, the Spanish Participation Group, University of Tokyo, University of Utah, Vanderbilt University, University of Virginia, University of Washington, and Yale University.




\bibliographystyle{mnras}

\begin{thebibliography}{}
\makeatletter
\relax
\def\mn@urlcharsother{\let\do\@makeother \do\$\do\&\do\#\do\^\do\_\do\%\do\~}
\def\mn@doi{\begingroup\mn@urlcharsother \@ifnextchar [ {\mn@doi@}
  {\mn@doi@[]}}
\def\mn@doi@[#1]#2{\def\@tempa{#1}\ifx\@tempa\@empty \href
  {http://dx.doi.org/#2} {doi:#2}\else \href {http://dx.doi.org/#2} {#1}\fi
  \endgroup}
\def\mn@eprint#1#2{\mn@eprint@#1:#2::\@nil}
\def\mn@eprint@arXiv#1{\href {http://arxiv.org/abs/#1} {{\tt arXiv:#1}}}
\def\mn@eprint@dblp#1{\href {http://dblp.uni-trier.de/rec/bibtex/#1.xml}
  {dblp:#1}}
\def\mn@eprint@#1:#2:#3:#4\@nil{\def\@tempa {#1}\def\@tempb {#2}\def\@tempc
  {#3}\ifx \@tempc \@empty \let \@tempc \@tempb \let \@tempb \@tempa \fi \ifx
  \@tempb \@empty \def\@tempb {arXiv}\fi \@ifundefined
  {mn@eprint@\@tempb}{\@tempb:\@tempc}{\expandafter \expandafter \csname
  mn@eprint@\@tempb\endcsname \expandafter{\@tempc}}}

\bibitem[\protect\citeauthoryear{{Adibekyan} et~al.,}{{Adibekyan}
  et~al.}{2013}]{afshsp13}
{Adibekyan} V.~Z.,  et~al., 2013, \mn@doi [\aap] {10.1051/0004-6361/201321520},
  \href {http://adsabs.harvard.edu/abs/2013A%26A...554A..44A} {554, A44}

\bibitem[\protect\citeauthoryear{{Allende Prieto}, {Beers}, {Wilhelm},
  {Newberg}, {Rockosi}, {Yanny}  \& {Lee}}{{Allende Prieto}
  et~al.}{2006}]{apbwnryl06}
{Allende Prieto} C.,  {Beers} T.~C.,  {Wilhelm} R.,  {Newberg} H.~J.,
  {Rockosi} C.~M.,  {Yanny} B.,   {Lee} Y.~S.,  2006, \mn@doi [\apj]
  {10.1086/498131}, \href {http://adsabs.harvard.edu/abs/2006ApJ...636..804A}
  {636, 804}

\bibitem[\protect\citeauthoryear{{Allende Prieto}, {Kawata}  \&
  {Cropper}}{{Allende Prieto} et~al.}{2016}]{apkc16}
{Allende Prieto} C.,  {Kawata} D.,   {Cropper} M.,  2016, \mn@doi [\aap]
  {10.1051/0004-6361/201629787}, \href
  {http://adsabs.harvard.edu/abs/2016A%26A...596A..98A} {596, A98}

\bibitem[\protect\citeauthoryear{{Aumer}, {Binney}  \& {Sch{\"o}nrich}}{{Aumer}
  et~al.}{2016}]{abs16}
{Aumer} M.,  {Binney} J.,   {Sch{\"o}nrich} R.,  2016, \mn@doi [\mnras]
  {10.1093/mnras/stw777}, \href
  {http://adsabs.harvard.edu/abs/2016MNRAS.459.3326A} {459, 3326}

\bibitem[\protect\citeauthoryear{{Bekki} \& {Tsujimoto}}{{Bekki} \&
  {Tsujimoto}}{2011}]{bt11}
{Bekki} K.,  {Tsujimoto} T.,  2011, \mn@doi [\apj] {10.1088/0004-637X/738/1/4},
  \href {http://adsabs.harvard.edu/abs/2011ApJ...738....4B} {738, 4}

\bibitem[\protect\citeauthoryear{{Bensby}, {Alves-Brito}, {Oey}, {Yong}  \&
  {Mel{\'e}ndez}}{{Bensby} et~al.}{2011}]{baboym11}
{Bensby} T.,  {Alves-Brito} A.,  {Oey} M.~S.,  {Yong} D.,   {Mel{\'e}ndez} J.,
  2011, \mn@doi [\apjl] {10.1088/2041-8205/735/2/L46}, \href
  {http://adsabs.harvard.edu/abs/2011ApJ...735L..46B} {735, L46}

\bibitem[\protect\citeauthoryear{{Bensby}, {Feltzing}  \& {Oey}}{{Bensby}
  et~al.}{2014}]{bfo14}
{Bensby} T.,  {Feltzing} S.,   {Oey} M.~S.,  2014, \mn@doi [\aap]
  {10.1051/0004-6361/201322631}, \href
  {http://ukads.nottingham.ac.uk/abs/2014A\%26A...562A..71B} {562, A71}

\bibitem[\protect\citeauthoryear{{Bergemann} et~al.,}{{Bergemann}
  et~al.}{2014}]{brsfa14}
{Bergemann} M.,  et~al., 2014, \mn@doi [\aap] {10.1051/0004-6361/201423456},
  \href {http://adsabs.harvard.edu/abs/2014A\%26A...565A..89B} {565, A89}

\bibitem[\protect\citeauthoryear{{Bird}, {Kazantzidis}  \& {Weinberg}}{{Bird}
  et~al.}{2012}]{jbkw12}
{Bird} J.~C.,  {Kazantzidis} S.,   {Weinberg} D.~H.,  2012, \mn@doi [\mnras]
  {10.1111/j.1365-2966.2011.19728.x}, \href
  {http://adsabs.harvard.edu/abs/2012MNRAS.420..913B} {420, 913}

\bibitem[\protect\citeauthoryear{{Bird}, {Kazantzidis}, {Weinberg}, {Guedes},
  {Callegari}, {Mayer}  \& {Madau}}{{Bird} et~al.}{2013}]{bkwgcmm13}
{Bird} J.~C.,  {Kazantzidis} S.,  {Weinberg} D.~H.,  {Guedes} J.,  {Callegari}
  S.,  {Mayer} L.,   {Madau} P.,  2013, \mn@doi [\apj]
  {10.1088/0004-637X/773/1/43}, \href
  {http://adsabs.harvard.edu/abs/2013ApJ...773...43B} {773, 43}

\bibitem[\protect\citeauthoryear{{Bland-Hawthorn} \&
  {Gerhard}}{{Bland-Hawthorn} \& {Gerhard}}{2016}]{bhg16}
{Bland-Hawthorn} J.,  {Gerhard} O.,  2016, \mn@doi [\araa]
  {10.1146/annurev-astro-081915-023441}, \href
  {http://adsabs.harvard.edu/abs/2016ARA%26A..54..529B} {54, 529}

\bibitem[\protect\citeauthoryear{{Bovy}, {Rix}, {Schlafly}, {Nidever},
  {Holtzman}, {Shetrone}  \& {Beers}}{{Bovy} et~al.}{2016}]{brsnhsb16}
{Bovy} J.,  {Rix} H.-W.,  {Schlafly} E.~F.,  {Nidever} D.~L.,  {Holtzman}
  J.~A.,  {Shetrone} M.,   {Beers} T.~C.,  2016, \mn@doi [\apj]
  {10.3847/0004-637X/823/1/30}, \href
  {http://adsabs.harvard.edu/abs/2016ApJ...823...30B} {823, 30}

\bibitem[\protect\citeauthoryear{{Brook}, {Kawata}, {Gibson}  \&
  {Freeman}}{{Brook} et~al.}{2004}]{bkgf04b}
{Brook} C.~B.,  {Kawata} D.,  {Gibson} B.~K.,   {Freeman} K.~C.,  2004, \mn@doi
  [\apj] {10.1086/422709}, \href
  {http://adsabs.harvard.edu/cgi-bin/nph-bib_query?bibcode=2004ApJ...612..894B&db_key=AST}
  {612, 894}

\bibitem[\protect\citeauthoryear{{Brook}, {Kawata}, {Martel}, {Gibson}  \&
  {Bailin}}{{Brook} et~al.}{2006}]{bkmg06}
{Brook} C.~B.,  {Kawata} D.,  {Martel} H.,  {Gibson} B.~K.,   {Bailin} J.,
  2006, \mn@doi [\apj] {10.1086/499154}, \href
  {http://adsabs.harvard.edu/cgi-bin/nph-bib_query?bibcode=2006ApJ...639..126B&db_key=AST}
  {639, 126}

\bibitem[\protect\citeauthoryear{{Brook} et~al.,}{{Brook}
  et~al.}{2012}]{bsgkh12}
{Brook} C.~B.,  et~al., 2012, \mn@doi [\mnras]
  {10.1111/j.1365-2966.2012.21738.x}, \href
  {http://adsabs.harvard.edu/abs/2012MNRAS.426..690B} {426, 690}

\bibitem[\protect\citeauthoryear{{Casagrande}, {Sch{\"o}nrich}, {Asplund},
  {Cassisi}, {Ram{\'{\i}}rez}, {Mel{\'e}ndez}, {Bensby}  \&
  {Feltzing}}{{Casagrande} et~al.}{2011}]{csacrmf11}
{Casagrande} L.,  {Sch{\"o}nrich} R.,  {Asplund} M.,  {Cassisi} S.,
  {Ram{\'{\i}}rez} I.,  {Mel{\'e}ndez} J.,  {Bensby} T.,   {Feltzing} S.,
  2011, \mn@doi [\aap] {10.1051/0004-6361/201016276}, \href
  {http://adsabs.harvard.edu/abs/2011A%26A...530A.138C} {530, A138}

\bibitem[\protect\citeauthoryear{{Chiappini} et~al.,}{{Chiappini}
  et~al.}{2015}]{carmm15}
{Chiappini} C.,  et~al., 2015, \mn@doi [\aap] {10.1051/0004-6361/201525865},
  \href {http://adsabs.harvard.edu/abs/2015A\%26A...576L..12C} {576, L12}

\bibitem[\protect\citeauthoryear{{Ciuc{\u a}}, {Kawata}, {Lin}, {Casagrande},
  {Seabroke}  \& {Cropper}}{{Ciuc{\u a}} et~al.}{2017}]{cklcsc17}
{Ciuc{\u a}} I.,  {Kawata} D.,  {Lin} J.,  {Casagrande} L.,  {Seabroke} G.,
  {Cropper} M.,  2017, preprint, \href
  {http://adsabs.harvard.edu/abs/2017arXiv170605005C} {} (\mn@eprint {arXiv}
  {1706.05005})

\bibitem[\protect\citeauthoryear{{Cresci}, {Mannucci}, {Maiolino}, {Marconi},
  {Gnerucci}  \& {Magrini}}{{Cresci} et~al.}{2010}]{cmmmgm10}
{Cresci} G.,  {Mannucci} F.,  {Maiolino} R.,  {Marconi} A.,  {Gnerucci} A.,
  {Magrini} L.,  2010, \mn@doi [\nat] {10.1038/nature09451}, \href
  {http://adsabs.harvard.edu/abs/2010Natur.467..811C} {467, 811}

\bibitem[\protect\citeauthoryear{{Curir}, {Lattanzi}, {Spagna}, {Matteucci},
  {Murante}, {Re Fiorentin}  \& {Spitoni}}{{Curir} et~al.}{2012}]{clsmmrfs12}
{Curir} A.,  {Lattanzi} M.~G.,  {Spagna} A.,  {Matteucci} F.,  {Murante} G.,
  {Re Fiorentin} P.,   {Spitoni} E.,  2012, \mn@doi [\aap]
  {10.1051/0004-6361/201118558}, \href
  {http://adsabs.harvard.edu/abs/2012A%26A...545A.133C} {545, A133}

\bibitem[\protect\citeauthoryear{{D'Onghia}, {Vogelsberger}  \&
  {Hernquist}}{{D'Onghia} et~al.}{2013}]{dvh13}
{D'Onghia} E.,  {Vogelsberger} M.,   {Hernquist} L.,  2013, \mn@doi [\apj]
  {10.1088/0004-637X/766/1/34}, \href
  {http://adsabs.harvard.edu/abs/2013ApJ...766...34D} {766, 34}

\bibitem[\protect\citeauthoryear{{Deason}, {Belokurov}, {Koposov}, {G{\'o}mez},
  {Grand}, {Marinacci}  \& {Pakmor}}{{Deason} et~al.}{2017}]{dbkggmp17}
{Deason} A.~J.,  {Belokurov} V.,  {Koposov} S.~E.,  {G{\'o}mez} F.~A.,  {Grand}
  R.~J.,  {Marinacci} F.,   {Pakmor} R.,  2017, \mn@doi [\mnras]
  {10.1093/mnras/stx1301}, \href
  {http://adsabs.harvard.edu/abs/2017MNRAS.470.1259D} {470, 1259}

\bibitem[\protect\citeauthoryear{{Feltzing}, {Bensby}  \&
  {Lundstr{\"o}m}}{{Feltzing} et~al.}{2003}]{fbl03}
{Feltzing} S.,  {Bensby} T.,   {Lundstr{\"o}m} I.,  2003, \mn@doi [\aap]
  {10.1051/0004-6361:20021661}, \href
  {http://adsabs.harvard.edu/abs/2003A\%26A...397L...1F} {397, L1}

\bibitem[\protect\citeauthoryear{{Foreman-Mackey}, {Hogg}, {Lang}  \&
  {Goodman}}{{Foreman-Mackey} et~al.}{2013}]{fmhlg13}
{Foreman-Mackey} D.,  {Hogg} D.~W.,  {Lang} D.,   {Goodman} J.,  2013, \mn@doi
  [\pasp] {10.1086/670067}, \href
  {http://adsabs.harvard.edu/abs/2013PASP..125..306F} {125, 306}

\bibitem[\protect\citeauthoryear{{Fuhrmann}}{{Fuhrmann}}{1998}]{kf98}
{Fuhrmann} K.,  1998, \aap, \href
  {http://ukads.nottingham.ac.uk/abs/1998A\%26A...338..161F} {338, 161}

\bibitem[\protect\citeauthoryear{{Gibson}, {Pilkington}, {Brook}, {Stinson}  \&
  {Bailin}}{{Gibson} et~al.}{2013}]{gpbsb13}
{Gibson} B.~K.,  {Pilkington} K.,  {Brook} C.~B.,  {Stinson} G.~S.,   {Bailin}
  J.,  2013, \mn@doi [\aap] {10.1051/0004-6361/201321239}, \href
  {http://adsabs.harvard.edu/abs/2013A\%26A...554A..47G} {554, A47}

\bibitem[\protect\citeauthoryear{{Gilmore} \& {Reid}}{{Gilmore} \&
  {Reid}}{1983}]{gilr83}
{Gilmore} G.,  {Reid} N.,  1983, \mnras, \href
  {http://ukads.nottingham.ac.uk/abs/1983MNRAS.202.1025G} {202, 1025}

\bibitem[\protect\citeauthoryear{{Grand}, {Kawata}  \& {Cropper}}{{Grand}
  et~al.}{2012}]{gkc12a}
{Grand} R.~J.~J.,  {Kawata} D.,   {Cropper} M.,  2012, \mn@doi [\mnras]
  {10.1111/j.1365-2966.2012.20411.x}, \href
  {http://adsabs.harvard.edu/abs/2012MNRAS.421.1529G} {421, 1529}

\bibitem[\protect\citeauthoryear{{Grand}, {Kawata}  \& {Cropper}}{{Grand}
  et~al.}{2014}]{gkc14}
{Grand} R.~J.~J.,  {Kawata} D.,   {Cropper} M.,  2014, \mn@doi [\mnras]
  {10.1093/mnras/stt2483}, \href
  {http://adsabs.harvard.edu/abs/2014MNRAS.439..623G} {439, 623}

\bibitem[\protect\citeauthoryear{{Grand}, {Kawata}  \& {Cropper}}{{Grand}
  et~al.}{2015}]{gkc15}
{Grand} R.~J.~J.,  {Kawata} D.,   {Cropper} M.,  2015, \mn@doi [\mnras]
  {10.1093/mnras/stv016}, \href
  {http://adsabs.harvard.edu/abs/2015MNRAS.447.4018G} {447, 4018}

\bibitem[\protect\citeauthoryear{{Grand}, {Springel}, {G{\'o}mez}, {Marinacci},
  {Pakmor}, {Campbell}  \& {Jenkins}}{{Grand} et~al.}{2016}]{gsgmp16}
{Grand} R.~J.~J.,  {Springel} V.,  {G{\'o}mez} F.~A.,  {Marinacci} F.,
  {Pakmor} R.,  {Campbell} D.~J.~R.,   {Jenkins} A.,  2016, \mn@doi [\mnras]
  {10.1093/mnras/stw601}, \href
  {http://adsabs.harvard.edu/abs/2016MNRAS.tmp..395G} {}

\bibitem[\protect\citeauthoryear{{Hayden} et~al.,}{{Hayden}
  et~al.}{2015}]{hbhnb15}
{Hayden} M.~R.,  et~al., 2015, \mn@doi [\apj] {10.1088/0004-637X/808/2/132},
  \href {http://ukads.nottingham.ac.uk/abs/2015ApJ...808..132H} {808, 132}

\bibitem[\protect\citeauthoryear{{Haywood}, {Di Matteo}, {Lehnert}, {Katz}  \&
  {G{\'o}mez}}{{Haywood} et~al.}{2013}]{hdmlkg13}
{Haywood} M.,  {Di Matteo} P.,  {Lehnert} M.~D.,  {Katz} D.,   {G{\'o}mez} A.,
  2013, \mn@doi [\aap] {10.1051/0004-6361/201321397}, \href
  {http://ukads.nottingham.ac.uk/abs/2013A\%26A...560A.109H} {560, A109}

\bibitem[\protect\citeauthoryear{{Jofr{\'e}} et~al.,}{{Jofr{\'e}}
  et~al.}{2016}]{jjvimhg16}
{Jofr{\'e}} P.,  et~al., 2016, \mn@doi [\aap] {10.1051/0004-6361/201629356},
  \href {http://adsabs.harvard.edu/abs/2016A%26A...595A..60J} {595, A60}

\bibitem[\protect\citeauthoryear{{Jones}, {Ellis}, {Richard}  \&
  {Jullo}}{{Jones} et~al.}{2013}]{jerj13}
{Jones} T.,  {Ellis} R.~S.,  {Richard} J.,   {Jullo} E.,  2013, \mn@doi [\apj]
  {10.1088/0004-637X/765/1/48}, \href
  {http://adsabs.harvard.edu/abs/2013ApJ...765...48J} {765, 48}

\bibitem[\protect\citeauthoryear{{Juri{\'c}} et~al.,}{{Juri{\'c}}
  et~al.}{2008}]{jibls08}
{Juri{\'c}} M.,  et~al., 2008, \mn@doi [\apj] {10.1086/523619}, \href
  {http://ukads.nottingham.ac.uk/abs/2008ApJ...673..864J} {673, 864}

\bibitem[\protect\citeauthoryear{{Kawata} \& {Chiappini}}{{Kawata} \&
  {Chiappini}}{2016}]{dkcc16}
{Kawata} D.,  {Chiappini} C.,  2016, \mn@doi [Astronomische Nachrichten]
  {10.1002/asna.201612421}, \href
  {http://adsabs.harvard.edu/abs/2016AN....337..976K} {337, 976}

\bibitem[\protect\citeauthoryear{{Kawata} \& {Gibson}}{{Kawata} \&
  {Gibson}}{2003}]{kg03a}
{Kawata} D.,  {Gibson} B.~K.,  2003, \mnras, \href
  {http://adsabs.harvard.edu/cgi-bin/nph-bib_query?bibcode=2003MNRAS.340..908K&db_key=AST}
  {340, 908}

\bibitem[\protect\citeauthoryear{{Kawata}, {Okamoto}, {Gibson}, {Barnes}  \&
  {Cen}}{{Kawata} et~al.}{2013}]{kogbc13}
{Kawata} D.,  {Okamoto} T.,  {Gibson} B.~K.,  {Barnes} D.~J.,   {Cen} R.,
  2013, \mn@doi [\mnras] {10.1093/mnras/sts161}, \href
  {http://adsabs.harvard.edu/abs/2013MNRAS.428.1968K} {428, 1968}

\bibitem[\protect\citeauthoryear{{Kawata}, {Grand}, {Gibson}, {Casagrande},
  {Hunt}  \& {Brook}}{{Kawata} et~al.}{2017}]{kggchb17}
{Kawata} D.,  {Grand} R.~J.~J.,  {Gibson} B.~K.,  {Casagrande} L.,  {Hunt}
  J.~A.~S.,   {Brook} C.~B.,  2017, \mn@doi [\mnras] {10.1093/mnras/stw2363},
  \href {http://adsabs.harvard.edu/abs/2017MNRAS.464..702K} {464, 702}

\bibitem[\protect\citeauthoryear{{Kordopatis}, {Wyse}, {Chiappini}, {Minchev},
  {Anders}  \& {Santiago}}{{Kordopatis} et~al.}{2017}]{kwcmas17}
{Kordopatis} G.,  {Wyse} R.~F.~G.,  {Chiappini} C.,  {Minchev} I.,  {Anders}
  F.,   {Santiago} B.,  2017, \mn@doi [\mnras] {10.1093/mnras/stx096}, \href
  {http://adsabs.harvard.edu/abs/2017MNRAS.tmp..161K} {}

\bibitem[\protect\citeauthoryear{{Kubryk}, {Prantzos}  \&
  {Athanassoula}}{{Kubryk} et~al.}{2013}]{kpa13}
{Kubryk} M.,  {Prantzos} N.,   {Athanassoula} E.,  2013, \mn@doi [\mnras]
  {10.1093/mnras/stt1667}, \href
  {http://adsabs.harvard.edu/abs/2013MNRAS.436.1479K} {436, 1479}

\bibitem[\protect\citeauthoryear{{Kubryk}, {Prantzos}  \&
  {Athanassoula}}{{Kubryk} et~al.}{2015}]{kpa15a}
{Kubryk} M.,  {Prantzos} N.,   {Athanassoula} E.,  2015, \mn@doi [\aap]
  {10.1051/0004-6361/201424171}, \href
  {http://adsabs.harvard.edu/abs/2015A%26A...580A.126K} {580, A126}

\bibitem[\protect\citeauthoryear{{Lee} et~al.,}{{Lee} et~al.}{2011}]{lbaij11}
{Lee} Y.~S.,  et~al., 2011, \mn@doi [\apj] {10.1088/0004-637X/738/2/187}, \href
  {http://adsabs.harvard.edu/abs/2011ApJ...738..187L} {738, 187}

\bibitem[\protect\citeauthoryear{{Leethochawalit}, {Jones}, {Ellis}, {Stark},
  {Richard}, {Zitrin}  \& {Auger}}{{Leethochawalit} et~al.}{2016}]{ljesrza16}
{Leethochawalit} N.,  {Jones} T.~A.,  {Ellis} R.~S.,  {Stark} D.~P.,  {Richard}
  J.,  {Zitrin} A.,   {Auger} M.,  2016, \mn@doi [\apj]
  {10.3847/0004-637X/820/2/84}, \href
  {http://adsabs.harvard.edu/abs/2016ApJ...820...84L} {820, 84}

\bibitem[\protect\citeauthoryear{{Loebman}, {Ro{\v s}kar}, {Debattista},
  {Ivezi{\'c}}, {Quinn}  \& {Wadsley}}{{Loebman} et~al.}{2011}]{lrdiqw11}
{Loebman} S.~R.,  {Ro{\v s}kar} R.,  {Debattista} V.~P.,  {Ivezi{\'c}} {\v Z}.,
   {Quinn} T.~R.,   {Wadsley} J.,  2011, \mn@doi [\apj]
  {10.1088/0004-637X/737/1/8}, \href
  {http://adsabs.harvard.edu/abs/2011ApJ...737....8L} {737, 8}

\bibitem[\protect\citeauthoryear{{Mackereth} et~al.,}{{Mackereth}
  et~al.}{2017}]{mbszcfgp17}
{Mackereth} J.~T.,  et~al., 2017, preprint, \href
  {http://adsabs.harvard.edu/abs/2017arXiv170600018M} {} (\mn@eprint {arXiv}
  {1706.00018})

\bibitem[\protect\citeauthoryear{{Martig} et~al.,}{{Martig}
  et~al.}{2015}]{mrahm15}
{Martig} M.,  et~al., 2015, \mn@doi [\mnras] {10.1093/mnras/stv1071}, \href
  {http://adsabs.harvard.edu/abs/2015MNRAS.451.2230M} {451, 2230}

\bibitem[\protect\citeauthoryear{{Masseron} \& {Gilmore}}{{Masseron} \&
  {Gilmore}}{2015}]{mg15}
{Masseron} T.,  {Gilmore} G.,  2015, \mn@doi [\mnras] {10.1093/mnras/stv1731},
  \href {http://adsabs.harvard.edu/abs/2015MNRAS.453.1855M} {453, 1855}

\bibitem[\protect\citeauthoryear{{McMillan}}{{McMillan}}{2011}]{pjm11}
{McMillan} P.~J.,  2011, \mn@doi [\mnras] {10.1111/j.1365-2966.2011.18564.x},
  \href {http://adsabs.harvard.edu/abs/2011MNRAS.414.2446M} {414, 2446}

\bibitem[\protect\citeauthoryear{{McMillan} \& {Dehnen}}{{McMillan} \&
  {Dehnen}}{2007}]{pjmwd07}
{McMillan} P.~J.,  {Dehnen} W.,  2007, \mn@doi [\mnras]
  {10.1111/j.1365-2966.2007.11753.x}, \href
  {http://adsabs.harvard.edu/abs/2007MNRAS.378..541M} {378, 541}

\bibitem[\protect\citeauthoryear{{Miglio} et~al.,}{{Miglio}
  et~al.}{2017}]{mcmdfgjk17}
{Miglio} A.,  et~al., 2017, preprint, \href
  {http://adsabs.harvard.edu/abs/2017arXiv170603778M} {} (\mn@eprint {arXiv}
  {1706.03778})

\bibitem[\protect\citeauthoryear{{Mikolaitis} et~al.,}{{Mikolaitis}
  et~al.}{2014}]{mhrbdv14}
{Mikolaitis} {\v S}.,  et~al., 2014, \mn@doi [\aap]
  {10.1051/0004-6361/201424093}, \href
  {http://adsabs.harvard.edu/abs/2014A\%26A...572A..33M} {572, A33}

\bibitem[\protect\citeauthoryear{{Minchev}, {Famaey}, {Quillen}, {Dehnen},
  {Martig}  \& {Siebert}}{{Minchev} et~al.}{2012}]{mfqdms12}
{Minchev} I.,  {Famaey} B.,  {Quillen} A.~C.,  {Dehnen} W.,  {Martig} M.,
  {Siebert} A.,  2012, \mn@doi [\aap] {10.1051/0004-6361/201219714}, \href
  {http://adsabs.harvard.edu/abs/2012A%26A...548A.127M} {548, A127}

\bibitem[\protect\citeauthoryear{{Minchev}, {Chiappini}  \& {Martig}}{{Minchev}
  et~al.}{2013}]{mcm13}
{Minchev} I.,  {Chiappini} C.,   {Martig} M.,  2013, \mn@doi [\aap]
  {10.1051/0004-6361/201220189}, \href
  {http://adsabs.harvard.edu/abs/2013A\%26A...558A...9M} {558, A9}

\bibitem[\protect\citeauthoryear{{Minchev}, {Chiappini}  \& {Martig}}{{Minchev}
  et~al.}{2014}]{mcm14}
{Minchev} I.,  {Chiappini} C.,   {Martig} M.,  2014, \mn@doi [\aap]
  {10.1051/0004-6361/201423487}, \href
  {http://adsabs.harvard.edu/abs/2014A\%26A...572A..92M} {572, A92}

\bibitem[\protect\citeauthoryear{{Minchev}, {Martig}, {Streich}, {Scannapieco},
  {de Jong}  \& {Steinmetz}}{{Minchev} et~al.}{2015}]{mmssdjs15}
{Minchev} I.,  {Martig} M.,  {Streich} D.,  {Scannapieco} C.,  {de Jong} R.~S.,
    {Steinmetz} M.,  2015, \mn@doi [\apjl] {10.1088/2041-8205/804/1/L9}, \href
  {http://adsabs.harvard.edu/abs/2015ApJ...804L...9M} {804, L9}

\bibitem[\protect\citeauthoryear{{Minchev}, {Steinmetz}, {Chiappini}, {Martig},
  {Anders}, {Matijevic}  \& {de Jong}}{{Minchev} et~al.}{2017}]{mscmamdj17}
{Minchev} I.,  {Steinmetz} M.,  {Chiappini} C.,  {Martig} M.,  {Anders} F.,
  {Matijevic} G.,   {de Jong} R.~S.,  2017, \mn@doi [\apj]
  {10.3847/1538-4357/834/1/27}, \href
  {http://adsabs.harvard.edu/abs/2017ApJ...834...27M} {834, 27}

\bibitem[\protect\citeauthoryear{{Navarro}, {Frenk}  \& {White}}{{Navarro}
  et~al.}{1997}]{nfw97}
{Navarro} J.~F.,  {Frenk} C.~S.,   {White} S.~D.~M.,  1997, \mn@doi [\apj]
  {10.1086/304888}, \href {http://adsabs.harvard.edu/abs/1997ApJ...490..493N}
  {490, 493}

\bibitem[\protect\citeauthoryear{{Nidever} et~al.,}{{Nidever}
  et~al.}{2014}]{nbbah14}
{Nidever} D.~L.,  et~al., 2014, \mn@doi [\apj] {10.1088/0004-637X/796/1/38},
  \href {http://adsabs.harvard.edu/abs/2014ApJ...796...38N} {796, 38}

\bibitem[\protect\citeauthoryear{{Petit}, {Krumholz}, {Goldbaum}  \&
  {Forbes}}{{Petit} et~al.}{2015}]{ptgf15}
{Petit} A.~C.,  {Krumholz} M.~R.,  {Goldbaum} N.~J.,   {Forbes} J.~C.,  2015,
  \mn@doi [\mnras] {10.1093/mnras/stv493}, \href
  {http://adsabs.harvard.edu/abs/2015MNRAS.449.2588P} {449, 2588}

\bibitem[\protect\citeauthoryear{{Price} \& {Monaghan}}{{Price} \&
  {Monaghan}}{2007}]{pm07}
{Price} D.~J.,  {Monaghan} J.~J.,  2007, \mn@doi [\mnras]
  {10.1111/j.1365-2966.2006.11241.x}, \href
  {http://ukads.nottingham.ac.uk/abs/2007MNRAS.374.1347P} {374, 1347}

\bibitem[\protect\citeauthoryear{{Prochaska}, {Naumov}, {Carney}, {McWilliam}
  \& {Wolfe}}{{Prochaska} et~al.}{2000}]{pncmw00}
{Prochaska} J.~X.,  {Naumov} S.~O.,  {Carney} B.~W.,  {McWilliam} A.,   {Wolfe}
  A.~M.,  2000, \mn@doi [\aj] {10.1086/316818}, \href
  {http://ukads.nottingham.ac.uk/abs/2000AJ....120.2513P} {120, 2513}

\bibitem[\protect\citeauthoryear{{Rahimi} \& {Kawata}}{{Rahimi} \&
  {Kawata}}{2012}]{rk12}
{Rahimi} A.,  {Kawata} D.,  2012, \mn@doi [\mnras]
  {10.1111/j.1365-2966.2012.20821.x}, \href
  {http://adsabs.harvard.edu/abs/2012MNRAS.422.2609R} {422, 2609}

\bibitem[\protect\citeauthoryear{{Recio-Blanco} et~al.,}{{Recio-Blanco}
  et~al.}{2014}]{rbdlkhhg14}
{Recio-Blanco} A.,  et~al., 2014, \mn@doi [\aap] {10.1051/0004-6361/201322944},
  \href {http://adsabs.harvard.edu/abs/2014A%26A...567A...5R} {567, A5}

\bibitem[\protect\citeauthoryear{{Reid} \& {Majewski}}{{Reid} \&
  {Majewski}}{1993}]{rm93}
{Reid} N.,  {Majewski} S.~R.,  1993, \mn@doi [\apj] {10.1086/172695}, \href
  {http://adsabs.harvard.edu/abs/1993ApJ...409..635R} {409, 635}

\bibitem[\protect\citeauthoryear{{Robin} \& {Creze}}{{Robin} \&
  {Creze}}{1986}]{rc86}
{Robin} A.,  {Creze} M.,  1986, \aap, \href
  {http://adsabs.harvard.edu/abs/1986A%26A...157...71R} {157, 71}

\bibitem[\protect\citeauthoryear{{Robin}, {Reyl{\'e}}, {Fliri}, {Czekaj},
  {Robert}  \& {Martins}}{{Robin} et~al.}{2014}]{rrfcrm14}
{Robin} A.~C.,  {Reyl{\'e}} C.,  {Fliri} J.,  {Czekaj} M.,  {Robert} C.~P.,
  {Martins} A.~M.~M.,  2014, \mn@doi [\aap] {10.1051/0004-6361/201423415},
  \href {http://ukads.nottingham.ac.uk/abs/2014A\%26A...569A..13R} {569, A13}

\bibitem[\protect\citeauthoryear{{Ruchti}, {Read}, {Feltzing}, {Pipino}  \&
  {Bensby}}{{Ruchti} et~al.}{2014}]{rrfpb14}
{Ruchti} G.~R.,  {Read} J.~I.,  {Feltzing} S.,  {Pipino} A.,   {Bensby} T.,
  2014, \mn@doi [\mnras] {10.1093/mnras/stu1435}, \href
  {http://adsabs.harvard.edu/abs/2014MNRAS.444..515R} {444, 515}

\bibitem[\protect\citeauthoryear{{Sch{\"o}nrich} \& {Binney}}{{Sch{\"o}nrich}
  \& {Binney}}{2009}]{sb09a}
{Sch{\"o}nrich} R.,  {Binney} J.,  2009, \mn@doi [\mnras]
  {10.1111/j.1365-2966.2009.14750.x}, \href
  {http://adsabs.harvard.edu/abs/2009MNRAS.396..203S} {396, 203}

\bibitem[\protect\citeauthoryear{{Sch{\"o}nrich} \& {McMillan}}{{Sch{\"o}nrich}
  \& {McMillan}}{2017}]{rspm17}
{Sch{\"o}nrich} R.,  {McMillan} P.~J.,  2017, \mn@doi [\mnras]
  {10.1093/mnras/stx093}, \href
  {http://adsabs.harvard.edu/abs/2017MNRAS.tmp..124S} {}

\bibitem[\protect\citeauthoryear{{Sellwood} \& {Binney}}{{Sellwood} \&
  {Binney}}{2002}]{jsjb02}
{Sellwood} J.~A.,  {Binney} J.~J.,  2002, \mn@doi [\mnras]
  {10.1046/j.1365-8711.2002.05806.x}, \href
  {http://adsabs.harvard.edu/abs/2002MNRAS.336..785S} {336, 785}

\bibitem[\protect\citeauthoryear{{Sellwood} \& {Carlberg}}{{Sellwood} \&
  {Carlberg}}{1984}]{jsrc84}
{Sellwood} J.~A.,  {Carlberg} R.~G.,  1984, \mn@doi [\apj] {10.1086/162176},
  \href {http://adsabs.harvard.edu/abs/1984ApJ...282...61S} {282, 61}

\bibitem[\protect\citeauthoryear{{Solway}, {Sellwood}  \&
  {Sch{\"o}nrich}}{{Solway} et~al.}{2012}]{sss12}
{Solway} M.,  {Sellwood} J.~A.,   {Sch{\"o}nrich} R.,  2012, \mn@doi [\mnras]
  {10.1111/j.1365-2966.2012.20712.x}, \href
  {http://ukads.nottingham.ac.uk/abs/2012MNRAS.422.1363S} {422, 1363}

\bibitem[\protect\citeauthoryear{{Spagna}, {Lattanzi}, {Re Fiorentin}  \&
  {Smart}}{{Spagna} et~al.}{2010}]{slrfs10}
{Spagna} A.,  {Lattanzi} M.~G.,  {Re Fiorentin} P.,   {Smart} R.~L.,  2010,
  \mn@doi [\aap] {10.1051/0004-6361/200913538}, \href
  {http://adsabs.harvard.edu/abs/2010A%26A...510L...4S} {510, L4}

\bibitem[\protect\citeauthoryear{{Stello} et~al.,}{{Stello}
  et~al.}{2015}]{shsjl15}
{Stello} D.,  et~al., 2015, \mn@doi [\apjl] {10.1088/2041-8205/809/1/L3}, \href
  {http://adsabs.harvard.edu/abs/2015ApJ...809L...3S} {809, L3}

\bibitem[\protect\citeauthoryear{{Stinson} et~al.,}{{Stinson}
  et~al.}{2013}]{sbrbr13}
{Stinson} G.~S.,  et~al., 2013, \mn@doi [\mnras] {10.1093/mnras/stt1600}, \href
  {http://adsabs.harvard.edu/abs/2013MNRAS.436..625S} {436, 625}

\bibitem[\protect\citeauthoryear{{Swinbank}, {Sobral}, {Smail}, {Geach},
  {Best}, {McCarthy}, {Crain}  \& {Theuns}}{{Swinbank}
  et~al.}{2012}]{ssgbmct12}
{Swinbank} A.~M.,  {Sobral} D.,  {Smail} I.,  {Geach} J.~E.,  {Best} P.~N.,
  {McCarthy} I.~G.,  {Crain} R.~A.,   {Theuns} T.,  2012, \mn@doi [\mnras]
  {10.1111/j.1365-2966.2012.21774.x}, \href
  {http://adsabs.harvard.edu/abs/2012MNRAS.426..935S} {426, 935}

\bibitem[\protect\citeauthoryear{{Toth} \& {Ostriker}}{{Toth} \&
  {Ostriker}}{1992}]{gtjpo92}
{Toth} G.,  {Ostriker} J.~P.,  1992, \mn@doi [\apj] {10.1086/171185}, \href
  {http://adsabs.harvard.edu/abs/1992ApJ...389....5T} {389, 5}

\bibitem[\protect\citeauthoryear{{Vera-Ciro}, {D'Onghia}, {Navarro}  \&
  {Abadi}}{{Vera-Ciro} et~al.}{2014}]{vcdna14}
{Vera-Ciro} C.,  {D'Onghia} E.,  {Navarro} J.,   {Abadi} M.,  2014, \mn@doi
  [\apj] {10.1088/0004-637X/794/2/173}, \href
  {http://adsabs.harvard.edu/abs/2014ApJ...794..173V} {794, 173}

\bibitem[\protect\citeauthoryear{{Wuyts} et~al.,}{{Wuyts}
  et~al.}{2016}]{wwffsg16}
{Wuyts} E.,  et~al., 2016, \mn@doi [\apj] {10.3847/0004-637X/827/1/74}, \href
  {http://adsabs.harvard.edu/abs/2016ApJ...827...74W} {827, 74}

\bibitem[\protect\citeauthoryear{{Xiang} et~al.,}{{Xiang}
  et~al.}{2015}]{xlyhw15}
{Xiang} M.-S.,  et~al., 2015, \mn@doi [Research in Astronomy and Astrophysics]
  {10.1088/1674-4527/15/8/009}, \href
  {http://adsabs.harvard.edu/abs/2015RAA....15.1209X} {15, 1209}

\bibitem[\protect\citeauthoryear{{Yamagata} \& {Yoshii}}{{Yamagata} \&
  {Yoshii}}{1992}]{yy92}
{Yamagata} T.,  {Yoshii} Y.,  1992, \mn@doi [\aj] {10.1086/116046}, \href
  {http://adsabs.harvard.edu/abs/1992AJ....103..117Y} {103, 117}

\bibitem[\protect\citeauthoryear{{Yang} \& {Krumholz}}{{Yang} \&
  {Krumholz}}{2012}]{ccymk12}
{Yang} C.-C.,  {Krumholz} M.,  2012, \mn@doi [\apj]
  {10.1088/0004-637X/758/1/48}, \href
  {http://adsabs.harvard.edu/abs/2012ApJ...758...48Y} {758, 48}

\bibitem[\protect\citeauthoryear{{Yong} et~al.,}{{Yong}
  et~al.}{2016}]{ycvckmm16}
{Yong} D.,  et~al., 2016, \mn@doi [\mnras] {10.1093/mnras/stw676}, \href
  {http://adsabs.harvard.edu/abs/2016MNRAS.459..487Y} {459, 487}

\bibitem[\protect\citeauthoryear{{Yoshii}}{{Yoshii}}{1982}]{yy82}
{Yoshii} Y.,  1982, \pasj, \href
  {http://ukads.nottingham.ac.uk/abs/1982PASJ...34..365Y} {34, 365}

\bibitem[\protect\citeauthoryear{{Yuan}, {Kewley}, {Swinbank}, {Richard}  \&
  {Livermore}}{{Yuan} et~al.}{2011}]{yksrl11}
{Yuan} T.-T.,  {Kewley} L.~J.,  {Swinbank} A.~M.,  {Richard} J.,   {Livermore}
  R.~C.,  2011, \mn@doi [\apjl] {10.1088/2041-8205/732/1/L14}, \href
  {http://ukads.nottingham.ac.uk/abs/2011ApJ...732L..14Y} {732, L14}

\makeatother
\end{thebibliography}







\bsp	
\label{lastpage}
\end{document}